\documentclass{aa}
\usepackage{graphicx,times}
\usepackage{bm,url,natbib}
\graphicspath{{./fig/}{./png/}}
\newcommand{\EQ}{\begin{equation}}
\newcommand{\EN}{\end{equation}}
\newcommand{\EQA}{\begin{eqnarray}}
\newcommand{\ENA}{\end{eqnarray}}
\newcommand{\eq}[1]{(\ref{#1})}

\newcommand{\Eq}[1]{Eq.~(\ref{#1})}

\newcommand{\Sec}[1]{Sect.~\ref{#1}}

\newcommand{\Fig}[1]{Fig.~\ref{#1}}

\newcommand{\Figs}[2]{Figs.~\ref{#1} and \ref{#2}}
\newcommand{\Figss}[2]{Figs.~\ref{#1}--\ref{#2}}
\newcommand{\Tab}[1]{Table~\ref{#1}}

\newcommand{\bra}[1]{\langle #1\rangle}

\newcommand{\meanrho}{\overline{\rho}}

{}
{}
{}

\newcommand{\meanSSSS}{\overline{\mbox{\boldmath ${\mathsf S}$}} {}}

{}
{}
{}
{}
{}
{}
{}
\newcommand{\meanAA}{\overline{\mbox{\boldmath $A$}}{}}{}
\newcommand{\meanBB}{\overline{\mbox{\boldmath $B$}}{}}{}
{}
{}
{}
{}
{}
{}
{}
\newcommand{\meanJJ}{\overline{\mbox{\boldmath $J$}}{}}{}
\newcommand{\meanUU}{\overline{\bm{U}}}

{}
{}
\newcommand{\meanQQ}{\overline{\mbox{\boldmath $Q$}}{}}{}

\newcommand{\meanB}{\overline{B}}

\newcommand{\meanU}{\overline{U}}

\newcommand{\meanp}{\overline{p}}

{}

{}
{}

\newcommand{\zzz}{\hat{\mbox{\boldmath $z$}} {}}

\newcommand{\nullvector}{{\bf0}}

\newcommand{\uu}{\mbox{\boldmath $u$} {}}
\newcommand{\UU}{\mbox{\boldmath $U$} {}}

\newcommand{\bb}{\mbox{\boldmath $b$} {}}
\newcommand{\BB}{\mbox{\boldmath $B$} {}}

\newcommand{\JJ}{\mbox{\boldmath $J$} {}}

\newcommand{\AAA}{\mbox{\boldmath $A$} {}}

\newcommand{\ff}{\mbox{\boldmath $f$} {}}

\newcommand{\FF}{\mbox{\boldmath $F$} {}}

\newcommand{\grav}{\mbox{\boldmath $g$} {}}
\newcommand{\nab}{\mbox{\boldmath $\nabla$} {}}
\newcommand{\OO}{\bm{\Omega}}

\newcommand{\SSSS}{\mbox{\boldmath ${\sf S}$} {}}

\newcommand{\DD}{{\rm D} {}}

\newcommand{\dd}{{\rm d} {}}

\newcommand{\const}{{\rm const}  {}}

\def\ga{\mathrel{\mathchoice {\vcenter{\offinterlineskip\halign{\hfil
$\displaystyle##$\hfil\cr>\cr\sim\cr}}}
{\vcenter{\offinterlineskip\halign{\hfil$\textstyle##$\hfil\cr>\cr\sim\cr}}}
{\vcenter{\offinterlineskip\halign{\hfil$\scriptstyle##$\hfil\cr>\cr\sim\cr}}}
{\vcenter{\offinterlineskip\halign{\hfil$\scriptscriptstyle##$\hfil\cr>\cr\sim\cr}}}}}
\def\Ma{\mbox{\rm Ma}}
\def\Co{\mbox{\rm Co}}

\def\Pm{\mbox{\rm Pr}_M}
\def\Rm{\mbox{\rm Re}_M}

\def\Rey{\mbox{\rm Re}}

\def\Co{\mbox{\rm Co}}

\def\tautd{\tau_{\rm td}}
\def\tauto{\tau_{\rm to}}
\def\betap{\beta_{\rm p}}
\def\betamin{\beta_{\rm min}}
\def\betastar{\beta_\star}
\def\csz{c_{\rm s0}}
\def\cs{c_{\rm s}}
\def\qp{q_{\rm p}}
\def\qpz{q_{\rm p0}}

\def\kf{k_{\rm f}}

\def\Peff{{\cal P}_{\rm eff}}
\def\Pmin{{\cal P}_{\rm eff}^{\rm min}}

\def\urms{u_{\rm rms}}

\def\nut{\nu_{\rm t}}
\def\etat{\eta_{\rm t}}
\def\etatz{\eta_{\rm t0}}

\def\Beq{B_{\rm eq}}
\def\Beqz{B_{\rm eq0}}

\def\half{{\textstyle{1\over2}}}

\def\onethird{{\textstyle{1\over3}}}

\newcommand{\G}{\,{\rm G}}

\newcommand{\kG}{\,{\rm kG}}

\newcommand{\s}{\,{\rm s}}

\newcommand{\Mm}{\,{\rm Mm}}

\newcommand{\yapj}[3]{ #1, {ApJ,} {#2}, #3}

\newcommand{\yapjl}[3]{ #1, {ApJL,} {#2}, #3}

\newcommand{\yan}[3]{ #1, {Astron.\ Nachr.,} {#2}, #3}

\newcommand{\yana}[3]{ #1, {A\&A,} {#2}, #3}

\newcommand{\yjfm}[3]{ #1, {J.\ Fluid Mech.,} {#2}, #3}

\newcommand{\ypfb}[3]{ #1, {Phys.\ Fluids B,} {#2}, #3}

\newcommand{\ysovl}[3]{ #1, {Sov.\ Astron.\ Lett.,} {#2}, #3}
\newcommand{\yjetp}[3]{ #1, {Sov.\ Phys.\ JETP,} {#2}, #3}

\newcommand{\ymn}[3]{ #1, {MNRAS,} {#2}, #3}

\newcommand{\ysph}[3]{ #1, {Solar Phys.,} {#2}, #3}

\newcommand{\ypre}[3]{ #1, {Phys.\ Rev.\ E,} {#2}, #3}

\newcommand{\yjcp}[3]{ #1, {J.\ Comput.\ Phys.,} {#2}, #3}
\newcommand{\yjour}[4]{ #1, {#2}, {#3}, #4}

\newcommand{\ybook}[3]{ #1, {#2} (#3)}

\title{Mean-field and direct numerical simulations of magnetic flux
concentrations from vertical field}

\authorrunning{A. Brandenburg et al.}
\author{A. Brandenburg\inst{1,2}, O. Gressel\inst{1,3}, S. Jabbari\inst{1,2},
N. Kleeorin\inst{4,1,5}, and I. Rogachevskii\inst{4,1,5}}

\institute{
Nordita, KTH Royal Institute of Technology and Stockholm University,
Roslagstullsbacken 23, SE-10691 Stockholm, Sweden
\and
Department of Astronomy, AlbaNova University Center,
Stockholm University, SE-10691 Stockholm, Sweden
\and
Niels Bohr International Academy, Niels Bohr Institute, Blegdamsvej 17, DK-2100,
Copenhagen \O, Denmark
\and
Department of Mechanical Engineering, Ben-Gurion University of the Negev,
POB 653, Beer-Sheva 84105, Israel
\and
Department of Radio Physics, N.~I.~Lobachevsky State University of
Nizhny Novgorod, Russia
}
\date{Received 16 September 2013 / Accepted 4 December 2013,~ $ $Revision: 1.248 $ $}
\begin{document}

\abstract{
Strongly stratified hydromagnetic turbulence has previously been found
to produce magnetic flux concentrations if the domain is large enough
compared with the size of turbulent eddies.
Mean-field simulations (MFS) using parameterizations of the Reynolds and Maxwell
stresses show a large-scale negative effective magnetic pressure instability and
have been able to reproduce many aspects of direct numerical simulations (DNS)
regarding growth rate, shape of
the resulting magnetic structures, and their height as a
function of magnetic field strength.
Unlike the case of an imposed horizontal field,
for a vertical one, magnetic flux concentrations of
equipartition strength with the turbulence can be reached,
resulting in magnetic spots that are reminiscent of sunspots.
}{
We determine under what conditions magnetic flux concentrations
with vertical field occur and what their internal structure is.
}{
We use a combination of MFS, DNS,
and implicit large-eddy simulations (ILES) to characterize the
resulting magnetic flux concentrations
in forced isothermal turbulence with an imposed vertical magnetic field.
}{
Using DNS, we confirm earlier results that in the kinematic stage of the
large-scale instability the horizontal wavelength of structures is about 10
times the density scale height.
At later times, even larger structures are being produced in a fashion
similar to inverse spectral transfer in helically driven turbulence.
Using ILES, we find that magnetic flux concentrations
occur for Mach numbers between 0.1 and 0.7.
They occur also for weaker stratification and larger turbulent eddies
if the domain is wide enough.
Using MFS, the size and aspect ratio of magnetic
structures are determined as functions of two input parameters
characterizing the parameterization of the effective magnetic pressure.
DNS, ILES, and MFS show magnetic flux tubes with mean-field energies
comparable to the turbulent kinetic energy.
These tubes can reach a length of about eight density scale heights.
Despite being $\le1\%$ equipartition strength, it is important that
their lower part is included within the computational domain to
achieve the full strength of the instability.
}{
The resulting vertical magnetic flux tubes
are being confined by downflows along the tubes and
corresponding inflow from the sides, which keep the field concentrated.
Application to sunspots remains a viable possibility.
}

\maketitle
\label{firstpage}

\section{Introduction}

Sunspots and active regions are generally thought to be the result
of magnetic fields emerging from deep at the bottom of the solar
convection zone \citep{Fan09}.
Alternatively, solar magnetic activity may be a shallow phenomenon \citep{B05}.
Several recent simulations with realistic physics of solar turbulent
convection with radiative transfer have demonstrated the appearance of
magnetic flux concentrations either spontaneously \citep{KKWM10,SN12}
or as a result of suitable initial conditions \citep{Cheung10,Rem11}.
There is also the phenomenon of magnetic flux expulsion, which has been
invoked as an explanation of the segregation of magneto-convection into
magnetized, non-convecting regions and non-magnetized, convecting ones
\citep{TWBP98}.

The magneto-hydrothermal structure of sunspots has been studied using
the thin flux tube approximation \citep{Spr81},
in which the stability and buoyant rise of magnetic fields
in the solar convection zone has been investigated.
This theory has been also applied to vertical magnetic
flux tubes, which open up toward the surface.
An important property of such tubes is the possibility of thermal collapse,
caused by an instability that leads to a downward shift of gas and a more
compressed magnetic field structure; see \cite{Spr79}, who adopted a
realistic equation of state including hydrogen ionization.
On the other hand, sunspot simulations of \cite{Rem11} and others
must make an ad hoc assumption about converging flows outside the
tube to prevent it from disintegrating due to turbulent convection.
This approach also does not capture the generation process,
that is now implicitly seen to operate in some of
the simulations of \cite{KKWM10} and \cite{SN12}.

To understand the universal physical mechanism
of magnetic flux concentrations, which has been argued to be
a minimal model of magnetic spot formation in the presence of
a vertical magnetic field \citep{BKR13},
we consider here forced turbulence
in a strongly stratified isothermal layer without radiation.
In the last few years, there has been significant progress in
modelling the physics of the resulting magnetic flux concentrations
in strongly stratified turbulence via the negative
effective magnetic pressure instability (NEMPI).
The physics behind this mechanism is the suppression of
total (hydrodynamic plus magnetic) turbulent pressure
by a large-scale magnetic field.
At large enough magnetic Reynolds numbers,
well above unity, the suppression of the total turbulent pressure
can be large, leading to a negative net effect.
In particular, the effective magnetic pressure
(the sum of non-turbulent and turbulent contributions)
becomes negative,
so that the large-scale negative effective magnetic pressure
instability is excited \citep{KRR89,KRR90,KMR93,KMR96,KR94,RK07}.

Hydromagnetic turbulence has been studied for decades \citep{Bis93},
but the effects of a large-scale magnetic field on the total pressure
are usually ignored, because in the incompressible case the pressure
can be eliminated from the problem.
This changes when there is gravitational density stratification,
even in the limit of small Mach number, because $\nab\cdot\rho\UU=0$
implies that $\nab\cdot\UU=U_z/H_\rho\neq0$.
Here, $\UU$ is the velocity, $H_\rho=|\dd\ln\rho/\dd z|^{-1}$ is the
density scale height, and gravity points in the negative $z$ direction.
When domain size and gravitational stratification are
big enough, the system can become unstable with respect to
NEMPI, which leads to a spontaneous accumulation of magnetic flux.
Direct numerical simulations (DNS)
with large scale separation have been used to verify this
mechanism for horizontal magnetic fields \citep{BKKMR11,KBKMR12,KBKMR13}.
In that case significant progress has been made in establishing the
connection between DNS and related mean-field simulations (MFS).
Both approaches show that the resulting magnetic flux concentrations are
advected downward in the nonlinear stage of NEMPI.
This is because the effective magnetic pressure is negative, so
when the magnetic field increases inside a horizontal flux structure,
gas pressure and density are locally increased to achieve pressure equilibrium,
thus making the effective magnetic buoyancy force negative.
This results in a downward flow (`potato-sack' effect).
Horizontal mean magnetic fields are advected downward by this flow and never
reach much more than a few percent of the equipartition field strength.

The situation is entirely different for vertical magnetic fields.
The downflow draws gas downward along magnetic field lines,
creating an underpressure in the upper parts, which concentrates
the magnetic field to equipartition field strength with respect
to the turbulent kinetic energy density \citep{BKR13}.
The resulting magnetic flux concentrations have superficially
the appearance of sunspots.
For horizontal fields, spots can also form and they have the
appearance of bipolar regions, as has been found in simulations
with a coronal layer above a turbulent region \citep{WLBKR13}.
However, to address the exciting possibility of explaining
the occurrence of sunspots by this mechanism, we need to know
more about the operation of NEMPI with a vertical magnetic field.
In particular, we need to understand how it is possible to
obtain magnetic field strengths much larger than the optimal
magnetic field strength at which NEMPI is excited.
We will do this through a detailed examination of
magnetic flux concentrations in MFS, where the origin of
flows can be determined unambiguously owing to the absence
of the much stronger turbulent convective motions.

We complement our studies with DNS and so-called `implicit large-eddy'
simulations (ILES), which are comparable to DNS in that they aim to
resolve the inertial range of the forced turbulence. ILES differ from
DNS in that one does not attempt to resolve the dissipation scale,
which is numerically expensive due to resolution requirements. In
short, ILES are DNS without explicit physical dissipation
coefficients.  However, unlike large-eddy simulations, no turbulence
parameterization model is used at all to represent the unresolved
scales. Lacking explicit dissipation, ILES instead rely on suitable
properties of the truncation error of the numerical scheme
\citep{2005IJNMF..47.1043G}, which guarantees that kinetic and
magnetic energies are dissipated near the grid scale.  In the
finite-volume code {\sc Nirvana} \citep{2004JCoPh.196..393Z} that we
use for ILES here, dissipation occurs in the averaging step of the
Godunov scheme.  The advantage of the finite-volume scheme is the
ability to capture shocks without explicit or artificial viscosity.
This allows us to probe the regime of higher Mach numbers without the
requirement to adjust the Reynolds number or grid resolution.

Following earlier work of \cite{BKKMR11}, we will stick to
the simple setup of an isothermal layer.
This is not only a computational convenience, but it is also
conceptually significant, because it allows us to disentangle
competing explanations for sunspot and active region formation.
One of them is the idea that active regions are being formed
and held in place by the more deeply rooted supergranulation
network at 20--40\,Mm depth \citep{SN12}.
In a realistic simulation there will be supergranulation and
large-scale downdrafts, but NEMPI also produces large-scale downdrafts
in the nonlinear stage of the evolution.
However, by using forced turbulence simulations in an isothermal layer,
an explanation in terms of supergranulation would not apply.

We emphasize that an isothermal layer can be infinitely extended.
Furthermore, the stratification is uniform in the sense that the
density scale height is independent of height.
Nevertheless, the density varies, so the equipartition magnetic field
strength also varies.
Therefore, the ratio of the imposed magnetic field strength to the
equipartition value varies with height.
NEMPI is excited at the height where this ratio is around 3\% \citep{LBKR13}.
This explains why NEMPI can be arranged to work at any field strength
if only the domain is tall enough.

At large domain size, DNS and ILES become expensive
and corresponding MFS are an ideal tool to address questions
concerning the global shape of magnetic flux concentrations.
In that case, significant conceptual simplifications can be
achieved by making use of the axisymmetry of the resulting
magnetic flux concentrations.
We also need to know more about the operation of NEMPI
under conditions closer to reality.
For example, how does it operate in the presence of larger gravity,
larger Mach numbers, and smaller scale separation?
This aspect is best being addressed through ILES,
where significant dissipation only occurs in shocks.

We consider three-dimensional (3-D) domains and compare in some
cases with MFS in two-dimensions (2-D) using axisymmetry or Cartesian geometry.
Here, axisymmetry is adequate for vertical tubes while Cartesian geometry
is adequate for vertical sheets of horizontal magnetic field.
The MFS provide guidance that is useful for understanding the
results of DNS and ILES,
so in this paper we begin with MFS, discuss
the mechanism of NEMPI and then focus
on the dependencies on gravity, scale separation,
and Mach numbers using DNS.
Finally, we assess the applicability of NEMPI to sunspot formation.

\section{Mean-field study of NEMPI}
\label{MFS}

For the analytical study of NEMPI with a vertical field we
consider the equations of mean-field MHD for mean
magnetic field $\meanBB$, mean velocity $\meanUU$,
and mean density $\meanrho$ in the anelastic approximation
for low Mach numbers, and for large fluid and magnetic Reynolds numbers,
\EQA
\label{dBmean}
{\partial\meanBB\over\partial t}&=& \nab \times \left(\meanUU\times\meanBB
-\etat\mu_0\meanJJ \right),
\\
\label{dUmean}
\meanrho{\DD\meanUU\over\DD t}&=&  -\nab \meanp_{\rm tot}
+ \mu_0^{-1} (\meanBB \cdot \nab)\meanBB
+\meanrho\grav - \nut \meanrho \meanQQ,
\\
\label{drhomean}
0&=&-\nab\cdot\meanrho \, \meanUU,
\ENA
where $\DD/\DD t=\partial/\partial t+\meanUU\cdot\nab$
is the advective derivative, $\meanp_{\rm tot}=
\meanp + p_{\rm eff}$ is the mean total
pressure, $\meanp$ is the mean gas pressure,
\EQ
p_{\rm eff}=(1-q_{\rm p})\meanBB^2/2\mu_0
\label{peff}
\EN
is the effective magnetic pressure \citep{KRR90,KMR93,KMR96},
$\meanrho$ is the mean density,
$\meanBB=\nab\times\meanAA+\zzz B_0$ is the mean
magnetic field with an imposed constant field
pointing in the $z$ direction,
$\meanJJ=\nab\times\meanBB/\mu_0$ is the mean
current density, $\mu_0$ is the vacuum permeability,
$\grav=(0,0,-g)$ is the
gravitational acceleration, $\etat$ is
the turbulent magnetic diffusivity, $\nut$ is the turbulent
viscosity,
\EQ
-\meanQQ=\nabla^2\meanUU+\onethird\nab\nab\cdot\meanUU
+2\meanSSSS\nab\ln\meanrho
\EN
is a term appearing in the viscous force with
\EQ
{\sf S}_{ij}=\half(\meanU_{i,j}+\meanU_{j,i})
-\onethird\delta_{ij}\nab\cdot\meanUU
\EN
being the traceless rate-of-strain tensor of the mean flow.

We adopt an isothermal equation of state with $p=\rho\cs^2$,
where $\cs=\const$ is the sound speed.
In the absence of a magnetic field, the hydrostatic equilibrium
solution is then given by $\meanrho=\rho_0\exp(-z/H_\rho)$,
where $H_\rho=\cs^2/g$ is the density scale height.

\subsection{Analytical estimates of growth rate of NEMPI}
\label{GrowthRate}

We linearize the mean-field equations~(\ref{dBmean})--(\ref{drhomean})
around the equilibrium:
$\meanUU_0=\bm{0}$, $\meanBB=\meanBB_0=\const$.
The equations for small perturbations (denoted by a tilde) can be
rewritten in the form
\begin{eqnarray}
&&{\partial \tilde {\bm B}\over \partial t} =
\nab \times \left(\tilde {\bm U}\times\meanBB_0 \right),
\label{C1}\\
&& \nab \cdot \tilde {\bm U} = {\tilde U_z\over
H_\rho},
\label{C2}\\
&&  {\partial \tilde {\bm U}\over \partial t} =
{1 \over \meanrho} \left[\mu_0^{-1}(\meanBB_0 \cdot
\nab)\tilde {\bm B}- \nab \tilde p_{\rm eff}
\right],
\label{C3}
 \end{eqnarray}
where
\begin{eqnarray}
\tilde p_{\rm tot}=\tilde p_{\rm eff} =
{2 \meanB_0 \, \tilde B_z \over \mu_0}
\left({\dd \Peff\over\dd\beta^2} \right)_{\beta=\beta_0}
\label{C4}
 \end{eqnarray}
with $\beta=\meanB/\Beq$ and $\Beq=\sqrt{\mu_0\meanrho}\urms$
is the local equipartition field strength, and $\urms$ is assumed to
be a constant in the present mean-field study.
Here, the effective magnetic pressure is written in normalized form as
\EQ
\Peff(\beta)\equiv \mu_0 p_{\rm eff}/\Beq^2= \half\left[1-\qp(\beta)\right]\beta^2.
\EN

In this section, we neglect dissipative terms such as the turbulent viscosity term
in the momentum equation and the turbulent magnetic
diffusion term in the induction equation.
We consider the axisymmetric problem,
use cylindrical coordinates $r, \varphi, z$ and introduce the magnetic
vector potential and stream function:
\begin{eqnarray}
\tilde {\bm B} = \nab {\times} \left(A {\rm e}_\varphi \right), \quad
\meanrho \, \tilde {\bm U} = \nab {\times}
\left(\Psi {\rm e}_\varphi \right).
\label{C5}
 \end{eqnarray}
Using the radial components of Eqs.~(\ref{C3}) and Eq.~(\ref{C1}) we
arrive at the following equation for the function
$\Phi(t,r,z) = \meanrho^{-1} \, \nabla_z \Psi$:
\begin{equation}
{\partial^2 \Phi\over \partial t^2}
=v^2_{\rm A}(z) \left[\nabla_z^2
+2 \left({\dd \Peff\over\dd\beta^2} \right)_{\beta=\beta_0}
\Delta_s \right] \Phi,
\label{C6}
\end{equation}
where $v_{\rm A}(z)=\meanB_0/\sqrt{\mu_0\meanrho(z)}$
is the mean Alfv\'en speed,
$\Delta_s$ is the radial part of the Stokes operator,
\begin{eqnarray*}
\Delta_s = {1 \over r} {\partial \over \partial r}
\left(r {\partial \over \partial r}\right) - {1 \over r^2},
\end{eqnarray*}
and we have used an exponential profile for the density stratification
in an isothermal layer,
\begin{eqnarray}
\meanrho=\meanrho_0 \exp (-z/H_\rho).
\label{meanrho_prof}
\end{eqnarray}
We seek solutions of Eq.~(\ref{C6}) in the form
\begin{eqnarray}
\Phi(t,r,z) = \exp (\lambda t)\, J_1(\sigma r / R) \, \Phi_0(z),
\label{C7}
 \end{eqnarray}
where $J_1(x)$ is the Bessel function of the first kind,
which satisfies the Bessel equation: $\Delta_s J_1(a r) = - a^2 J_1(a r)$.
Substituting Eq.~(\ref{C7}) into Eq.~(\ref{C6}), we obtain
the equation for the function $\Phi_0(z)$:
\begin{equation}
{d^2\Phi_0 \over dz^2} - \left[{\lambda^2 \over v^2_{\rm A}(z)}
+{2\sigma^2 \over R^2} \left({\dd \Peff\over\dd\beta^2} \right)_{\beta=\beta_0}
\right] \Phi_0 =0.
\label{C8}
\end{equation}
For $R^2 \Phi''_0(z)/\Phi_0 \ll 1$, the growth rate of NEMPI
is given by
\begin{eqnarray}
\lambda={v_{\rm A} \sigma \over R} \left[
-2\left(\dd \Peff \over \dd\beta^2 \right)_{\beta=\beta_0}\right]^{1/2} .
\label{C10}
\end{eqnarray}
This equation shows that, compared to the case of a horizontal
magnetic field, where there was a factor $H_\rho$ in the denominator,
in the case of a vertical field the relevant length is $R/\sigma$.
Introducing as a new variable $X=\beta_0^2(z)$, we can rewrite
Eq.~(\ref{C8}) in the form
\begin{equation}
X^3 {d^2\Phi_0 \over dX^2} + X^2 {d\Phi_0 \over dX}
- \left({\lambda^2 H_\rho^2\over \urms^2}
+{2\sigma^2 H_\rho^2\over R^2} X {d \Peff\over dX}
\right) \Phi_0 =0.\quad
\label{C9}
\end{equation}
We now need to make detailed assumptions about the functional form
of $\Peff(\beta^2)$.
A useful parameterization of $\qp$ in \Eq{peff} is \citep{KBKR12}
\EQ
\qp={\qpz\over1+\beta^2/\betap^2}\equiv{\betastar^2\over\betap^2+\beta^2},
\EN
where $\betastar=\sqrt{\qpz}\betap$.
It is customary to obtain approximate analytic solutions to Eq.~(\ref{C9})
as marginally bound states of the associated Schr\"odinger equation,
$\Psi_0''-\tilde{U}(X) \, \Psi_0=0$, via the transformation
$\Phi_0=\Psi_0 /\sqrt{X}$, where
\begin{eqnarray}
\tilde{U}(X) = {\lambda^2 H_\rho^2\over \urms^2 X^3} - {1 \over 4X^2}
+{\sigma^2 H_\rho^2\over R^2 X^2}\left(1 - {\qpz \over (1+X^2/\betap^2)^2}\right)
,
\nonumber\\
\label{C11}
\end{eqnarray}
where primes denote a derivative with respect to $X$.
The potential $\tilde{U}(X)$ has the following asymptotic behavior:
$\tilde U \to \lambda^2 H_\rho^2/ (\urms^2 X^3)$ for small $X$, and
$\tilde U(X) \to (\sigma^2 H_\rho^2/ R^2 -1/4) X^{-2}$ for large $X$.
For the existence of an instability, the potential $\tilde U(R)$ should
have a negative minimum.
However, the exact values of the growth rate of NEMPI, the scale at which
the growth rate attains the maximum value, and
how the resulting magnetic field structure looks like in the nonlinear
saturated regime of NEMPI can only be obtained numerically using MFS.

\subsection{MFS models}

For consistency with earlier studies,
we keep the governing MFS parameters equal to those used in a
recent study by \cite{LBKR13}.
Thus, unless stated otherwise, we use the values
\EQ
\qpz=32,\quad\betap=0.058\quad\mbox{(reference model)},
\label{qpz}
\EN
which are based on Eq.~(22) of \cite{BKKR12},
applied to $\Rm=18$.

The mean-field equations are solved numerically without making
the anelastic approximation, i.e., we solve
\EQ
{\partial\meanrho \over\partial t}=-\nab\cdot\meanrho\meanUU
\label{drhomeandt}
\EN
together with the equations for the mean
vector potential $\meanAA$ such that
$\meanBB=\BB_0+\nab\times\meanAA$
is divergence-free, the mean velocity $\meanUU$, and the
mean density $\meanrho$, using the
{\sc Pencil Code} (\url{http://pencil-code.googlecode.com}),
which has a mean-field module built in and is used for calculations
both in Cartesian and cylindrical geometries.
Here, $\BB_0=(0,0,B_0)$ is the imposed uniform vertical field.
The respective coordinate systems are $(x,y,z)$ and $(r,\varphi,z)$.
In the former case we use periodic boundary conditions in the
horizontal directions, $-L_\perp/2<(x,y)<L_\perp/2$,
while in the latter we adopt perfect conductor,
free-slip boundary conditions at the side walls at $r=L_r$
and regularity conditions on the axis.
On the upper and lower boundaries at $z=z_{\rm top}$ and $z=z_{\rm bot}$
we use in both geometries stress-free conditions,
$\zzz\times\partial\meanUU/\partial z=\nullvector$ and $\zzz\cdot\meanUU=0$,
and assume the magnetic field to be normal to the boundary,
i.e., $\zzz\times\meanBB=\nullvector$.

Following earlier work, we display results for the magnetic field either
by normalizing with $B_0$, which is a constant, or
by normalizing with $\Beq$, which decreases with height.
The strength of the imposed field is often specified in terms of
$\Beqz=\Beq(z=0)$.

\subsection{Nondimensionalization}
\label{Nondimensionalization}

Nondimensional parameters are indicated by tildes and hats, and include
$\tilde{B}_0=B_0/(\mu_0\rho_0\cs^2)^{1/2}$
and $\tilde\etat=\etat/\cs H_\rho$, in addition
to parameters in \Eq{qpz} characterizing the functional form of $\qp(\beta)$.
Additional quantities include
$\tilde\kf=\kf H_\rho$ and $\hat\kf=\kf/k_1$,
where a hat is used to indicate nondimensionalization that uses
quantities other than $\cs$ and $H_\rho$, such as $k_1=2\pi/L_\perp$,
which is the lowest horizontal
wavenumber in a domain with horizontal extent $L_\perp$.
For example, $\hat{g}=g/\cs^2 k_1$, is nondimensional gravity and
$\hat{\lambda}=\lambda H_\rho^2/\etat$ is the nondimensional growth rate.
It is convenient to quote also $B_0/\Beqz$ with $\Beqz=\Beq(z=0)$.
Note that $B_0/\Beqz$ is larger than $\tilde{B}_0$ by the inverse
of the turbulent Mach number, $\Ma=\urms/\cs$.
It is convenient to normalize the mean flow by $\urms$ and denote it
by a hat, i.e., $\hat{\meanU}=\meanU/\urms$.
Likewise, we define $\hat{\meanB}=\meanB/\Beq$.

In MFS, the value of $\etat$ is assumed to be given by $\etatz=\urms/3\kf$.
Using the test-field method, \cite{SBS08} found this to be an accurate
approximation of $\etat$.
Thus, we have to specify both $\Ma$ and $\tilde\kf$.
In most of our runs we use $\Ma=0.1$ and $\tilde\kf=33$,
corresponding to $\tilde\etat=10^{-3}$.
Furthermore, $\kf$ and $H_\rho$ are in principle not independent
of each other either.
In fact, mixing length theory suggests $\kf H_\rho\approx6.5$
\citep{LBKR13}, but it would certainly be worthwhile to
compute this quantity from high-resolution convection simulations
spanning multiple scale heights.
However, in this paper, different values of $\kf H_\rho$ are
considered.
With these preparations in place, we can now address questions
concerning the horizontal wavelength of the instability and the
vertical structure of the magnetic flux tubes.

\begin{figure}\begin{center}
\includegraphics[width=\columnwidth]{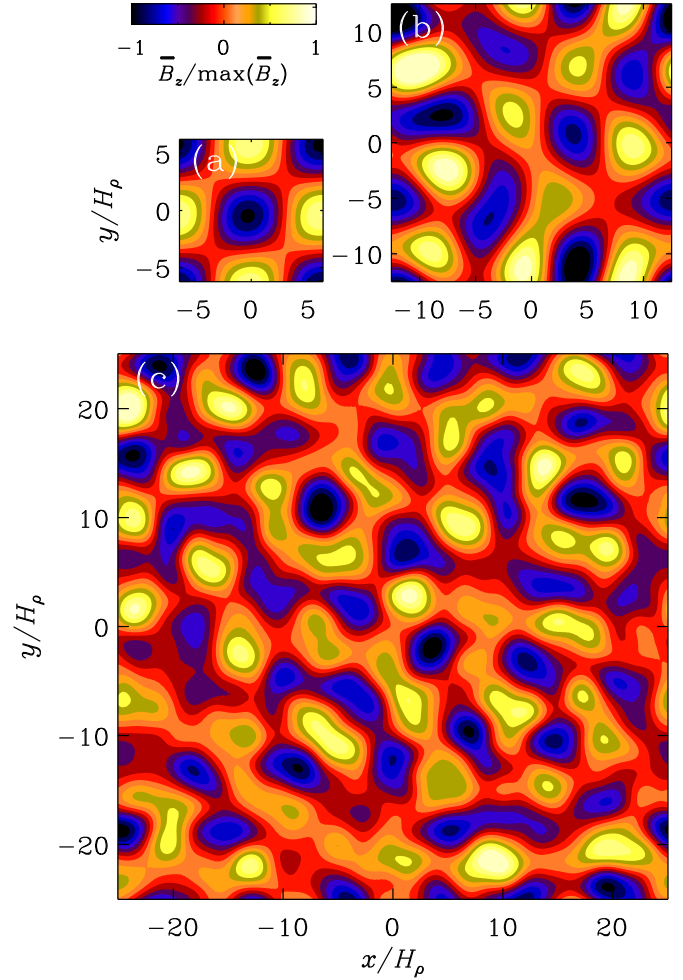}
\end{center}\caption[]{
Horizontal patterns of $\meanB_z$ at $z=0$ from a 3-D MFS during the
kinematic growth phase with $B_0/\Beqz=0.1$ and horizontal extents
with (a) $L_\perp/H_\rho=4\pi$, (b) $8\pi$, and (c) $16\pi$.
}\label{pBz_xy_mid_all}
\end{figure}

\begin{figure}\begin{center}
\includegraphics[width=\columnwidth]{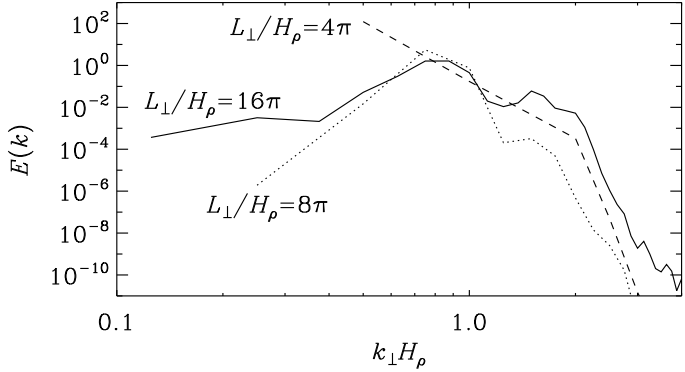}
\end{center}\caption[]{
Power spectra of $\meanB_z$ for different horizontal domain sizes
at $z=0$ from a 3-D MFS during the kinematic growth phase with $B_0/\Beqz=0.1$.
}\label{pBz_xy_mid_all_power}
\end{figure}

\subsection{Aspect ratio of NEMPI}
\label{AspectRatio}

The only natural length scale in an isothermal layer
in MFS is $H_\rho$.
It determines the scale of NEMPI.
At onset, the horizontal scale of the magnetic field
pattern will be a certain multiple of $H_\rho$.
In the following we denote the corresponding horizontal wavenumber
of this pattern by $k_\perp$.
Earlier work by \cite{KBKMR13} showed that for an imposed horizontal
magnetic field we have $k_\perp H_\rho\approx0.8$...$1$.
This pattern was 2-D in the plane perpendicular
to the direction of the imposed magnetic field, corresponding to
horizontal rolls oriented along the mean magnetic field.
In the present case of a vertical field, the magnetic perturbations
have a cellular pattern with horizontal wavenumber $k_\perp$.
To determine the value of $k_\perp H_\rho$ for the
case of an imposed vertical
magnetic field, we have to ensure that the number of cells
per unit area is independent of the size of the domain.
In \Fig{pBz_xy_mid_all}, we compare MFS with horizontal
aspect ratios ranging from 2 to 8.
We see that the magnetic pattern is fully captured in a domain with
normalized horizontal extent $L_\perp/H_\rho=4\pi$, i.e., the horizontal scale
of the magnetic field pattern is twice the value of $H_\rho$,
i.e., $k_x=k_y=H_\rho^{-1}/2$, so that
$k_\perp \equiv (k_x^2+k_y^2)^{1/2}=H_\rho^{-1}/\sqrt{2}$,
or $k_\perp H_\rho\approx0.7$.
The value $k_\perp H_\rho\approx0.7$ is also confirmed by taking
a power spectrum of $\meanB_z(x,y)$; see \Fig{pBz_xy_mid_all_power},
which shows a peak at a similar value.

Comparing the three simulations shown in \Fig{pBz_xy_mid_all}, we see that
a regular checkerboard pattern is only obtained for the smallest domain size;
see \Fig{pBz_xy_mid_all}(a).
For larger domain sizes the patterns are always irregular such
that a cell of one sign can be surrounded by 3--5 cells of the
opposite sign.
Nevertheless, in all three cases we have approximately the same number
of cells per unit area.

\begin{figure}\begin{center}
\includegraphics[width=\columnwidth]{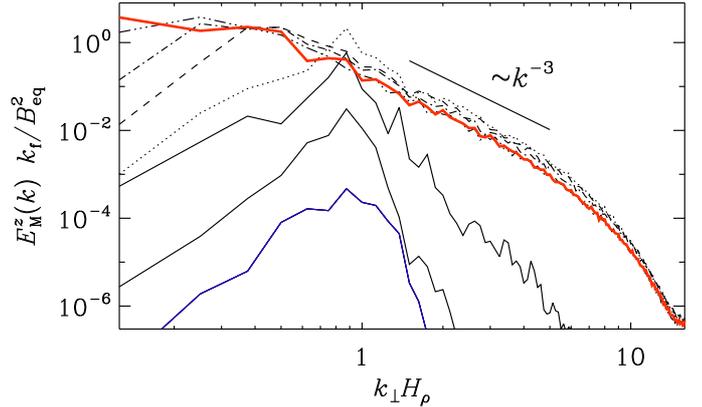}
\end{center}\caption[]{
Time evolution of normalized spectra of $B_z$
from 3-D MFS during the late nonlinear phase
at the top of the domain, $k_1 z=\pi$, at normalized times
$t\etatz/H_\rho^2\approx5$ (blue), 6, 7, 10, 20, 30, 40, and 50 (red),
with $g=\cs^2 k_1$,
$B_0/\Beqz=0.1$, and $L_\perp/H_\rho=16\pi$
}\label{pBz_spec2_256x256x128_16pi_B01a}
\end{figure}

In the nonlinear regime, structures continue to merge and more power
is transferred to lower horizontal wavenumbers;
see \Fig{pBz_spec2_256x256x128_16pi_B01a}.
Later in \Sec{DependenceScaleSeparation} we present similar results
also for our DNS.

\subsection{Vertical magnetic field profile during saturation}
\label{VerticalProfile}

In an isothermal atmosphere, the scale height is constant and there is no
physical upper boundary, so we can extend the computation in the $z$ direction
at will, although the magnetic pressure will strongly exceed the
turbulent pressure at large heights, which can pose computational difficulties.
To study the full extent of magnetic flux concentrations, we need a
big enough domain.
In the following we consider the range $-3\pi\leq z/H_\rho\leq3\pi$,
which results in a density contrast of more than $10^8$.
To simplify matters, we restrict ourselves
in the present study
to axisymmetric calculations which are faster than
3-D Cartesian ones.

\begin{figure}\begin{center}
\includegraphics[width=\columnwidth]{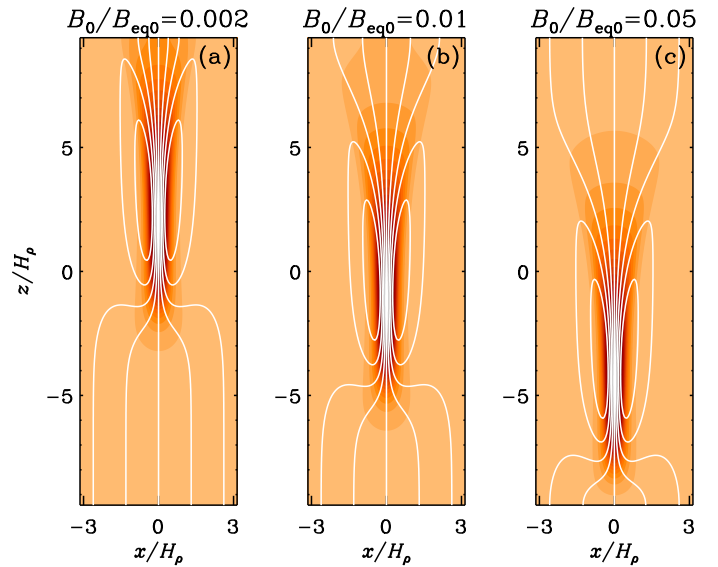}
\end{center}\caption[]{
Comparison of magnetic field profiles from axisymmetric MFS
for Runs~Bv002/33--Bv05/33 with three values of $B_0/\Beqz$
and $\tilde\etat=10^{-3}$, corresponding to $\kf H_\rho=33$.
}\label{ppfline_comp}
\end{figure}

\begin{figure}\begin{center}
\includegraphics[width=.84\columnwidth]{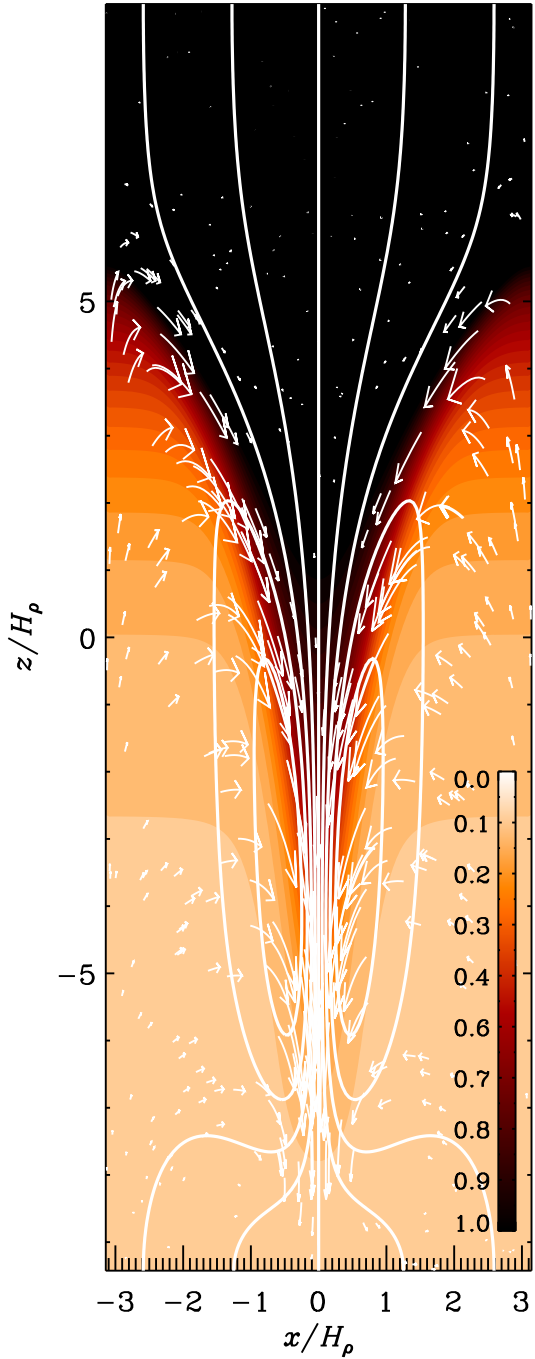}
\end{center}\caption[]{
$\meanB_z/\Beq$ together with field lines and flow vectors from MFS,
for Run~Bv05/33 with $B_0/\Beqz=0.05$.
The flow speed varies from $-0.27\urms$ (downward) to $0.08\urms$ (upward).
}\label{ppfline_flow}
\end{figure}

\begin{figure}\begin{center}
\includegraphics[width=\columnwidth]{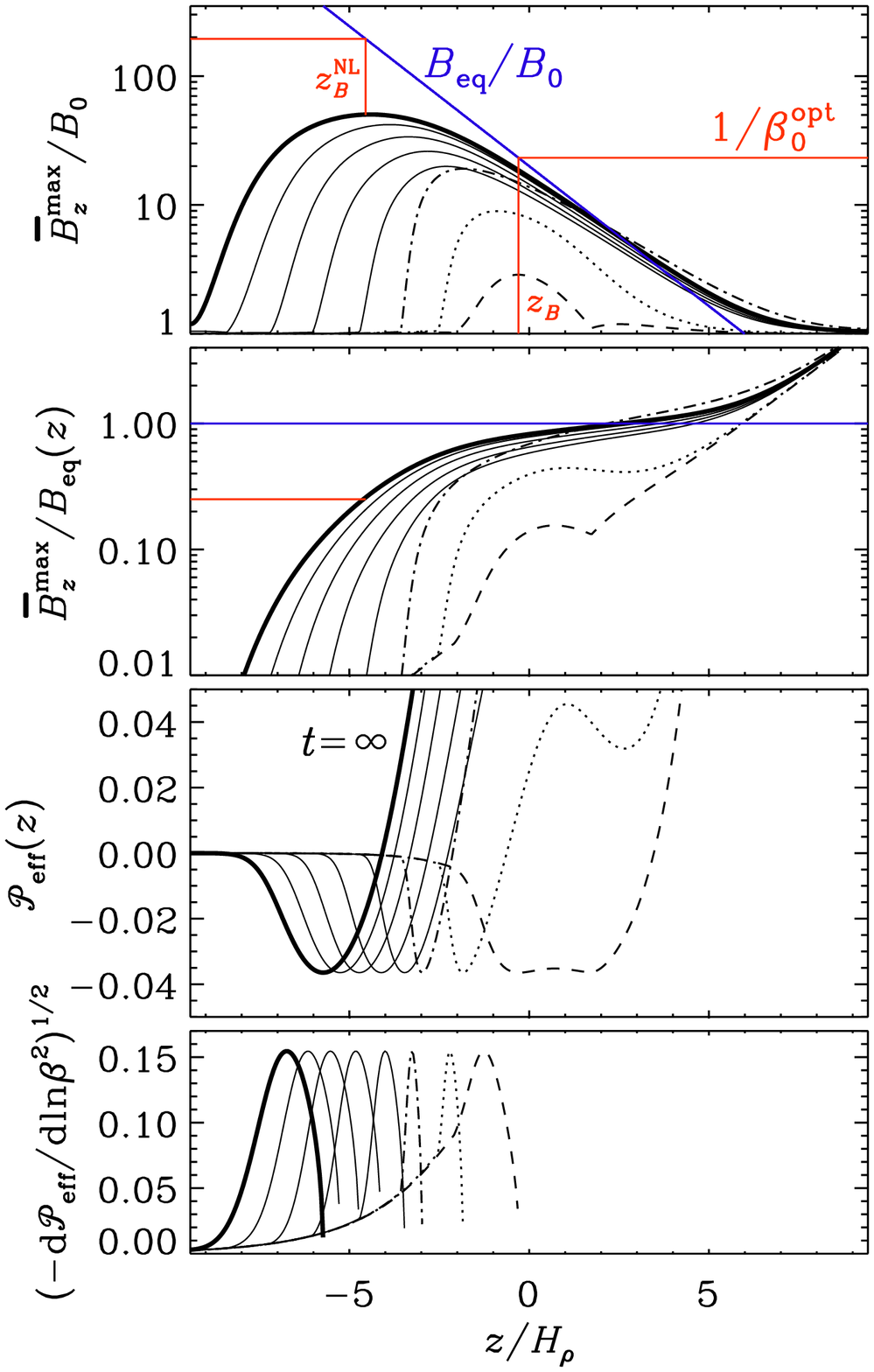}
\end{center}\caption[]{
Time evolution of normalized vertical magnetic field profiles,
(a) $\meanB_z^{\rm max}/B_0$ together with
$\Beq(z)/B_0$ (shown by blue line), (b) $\meanB_z^{\rm max}/\Beq(z)$, as well as
(c) $\Peff(z)$ and (d) $(-\dd\Peff/\dd\ln\beta^2)^{1/2}$,
from a MFS for Run~Bv05/33 with $B_0/\Beqz=0.05$
at $t/\tautd=2.9$ (dashed), 3 (dotted), 3.1 (dash-dotted),
3.3, 3.7, 4.2., 5, and 50 (thick solid line).
The blue solid lines indicate $\Beq(z)$, normalized by (a) $B_0$ and
(b) by itself (corresponding thus to unity).
The red lines indicate the locations $z_B$ and $z_B^{\rm NL}$,
as well as relevant intersections with normalized values of
$\meanB_z^{\rm max}$ and $\Beq$.
}\label{pBmax_axi}
\end{figure}

In \Fig{ppfline_comp} we compare the results
for the mean magnetic field profiles
for three values of $B_0/\Beqz$ ranging from 0.002 to 0.05.
These values are smaller than those studied in \Sec{AspectRatio},
because in the nonlinear regime and in a deeper domain the structures
are allowed to sink by a substantial amount.
By choosing $B_0/\Beqz$ to be smaller, the tubes are fully
contained in our domain.
As $B_0$ increases, we expect the position of the magnetic flux tube,
$z_B$, to move downward like
\EQ
z_B=z_{B0}-2H_\rho\ln(B_0/\Beqz),
\label{zB}
\EN
where $z_{B0}=2H_\rho\ln\beta_0^{\rm opt}$ is a reference height and
\EQ
\beta_0^{\rm opt}\equiv B_0/\Beq(z_B)\approx0.03...0.06
\label{zBcondition}
\EN
is the optimal normalized field strength for NEMPI to be excited \citep{LBKR13}.
The validity of \Eq{zB} can be verified through \Fig{ppfline_comp},
where $B_0/\Beqz$ increases by
a factor of 25, corresponding to $\Delta z_B=-6.4$.

In all cases, we obtain a slender tube with approximate aspect ratio of 1:8.
In other words, the shape of the magnetic field lines is the same
for all three values of $B_0/\Beqz$,
and just the position of the magnetic flux concentration shifts in the
vertical direction.
Note in particular that the thickness of structures is always the same.
This is different from the nonlinear MFS in Cartesian geometry discussed
above, where structures are able to merge.
Merging is not really possible in the same way in an axisymmetric
container, because any additional structure would correspond to a ring.

The mean flow structure associated with the magnetic flux tube
is shown in \Fig{ppfline_flow} for Run~Bv05/33 with $B_0/\Beqz=0.05$.
We find inflow into the tube along field lines at large heights and outflow
at larger depth.
The vertical component of the flow in the tube points always downward,
i.e., there is no obvious effect from positive magnetic buoyancy.
The maximum downflow speed is about $0.27\urms$, so it is subdominant
compared with the turbulent velocity, but this could be enough to cause
a noticeable temperature change in situations where the energy equation
is solved.

The resulting magnetic field lines look roughly similar to those of the DNS
with an imposed vertical magnetic field \citep{BKR13}.
In DNS, however, the thickness of the magnetic flux tube is larger
than in the MFS by about a factor of three.
This discrepancy could be explained if the actual value of $\etat$
was in fact larger than the estimate given by $\etatz$.
We return to this possibility in \Sec{DependenceScaleSeparation}.
Alternatively, it might be related to the possibility that the
coefficients in \Eq{qpz} could actually be different.

The time evolution of the vertical magnetic field profiles,
$\meanB_z^{\rm max}/B_0$ and $\meanB_z^{\rm max}/\Beq(z)$, is shown
in \Fig{pBmax_axi} at different times for the case $B_0/\Beqz=0.05$,
corresponding to \Fig{ppfline_comp}(c).
Here, we also show the time evolution of the corresponding profiles
of $\Peff(z)$ and $(-\dd\Peff/\dd\ln\beta^2)^{1/2}$.
In the kinematic regime, the peak of the latter quantity is a good indicator
of the peak of the eigenfunction \citep{KBKMR13}.
In the present case, the magnetic field in the kinematic phase
peaks at a height $z_B$ that is given by the condition \eq{zBcondition}.
According to the MFS of \cite{LBKR13},
this condition is approximately the same for vertical and horizontal fields.
Looking at \Fig{pBmax_axi} for $B_0/\Beqz=0.05$,
we see that at $z/H_\rho\approx-0.5$ we
have $\Beq/B_0\approx33$, which agrees with \Eq{zBcondition}.
However, unlike the case of a horizontal magnetic field, where in
the kinematic phase the mean field was found to peak at a height
below that where $(-\dd\Peff/\dd\ln\beta^2)^{1/2}$ peaks,
we now see that the field peaks above that position.

As NEMPI begins to saturate, the peak of $\meanB_z^{\max}$
moves further down to $z=z_B^{\rm NL}\approx-5\,H_\rho$ during the next one or two
turbulent diffusive times.
By that time, $\meanB_z^{\max}$ has reached values up to
$\meanB_z^{\max}/B_0\approx50$.
At that depth, $\meanB_z^{\max}/\Beq(z)$ is about 0.25,
but this quantity continues to increase with height
and reaches super-equipartition values at $z/H_\rho\approx3$
(second panel of \Fig{pBmax_axi}).

\subsection{Smaller scale separation}

\begin{figure}\begin{center}
\includegraphics[width=\columnwidth]{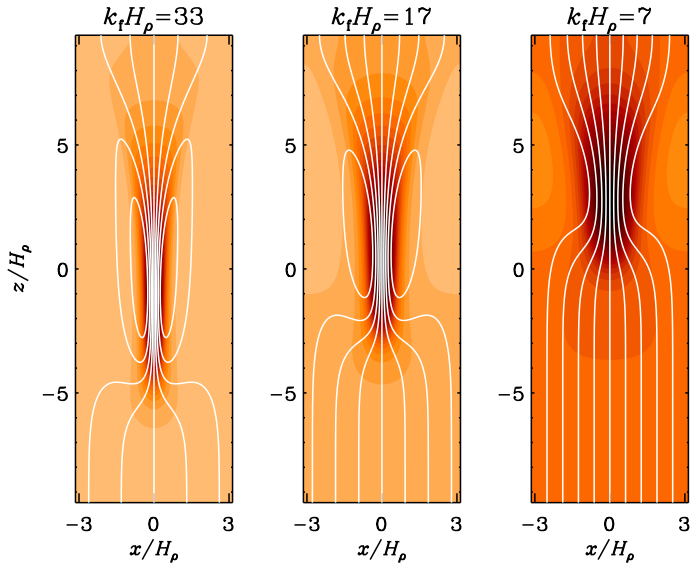}
\end{center}\caption[]{
Comparison of magnetic field profiles
from an axisymmetric MFS for Runs~Bv01/33--Bv01/7
with $B_0/\Beqz=0.01$ and three values of $\kf H_\rho$.
}\label{ppfline_comp_etat}
\end{figure}

\begin{figure}\begin{center}
\includegraphics[width=.94\columnwidth]{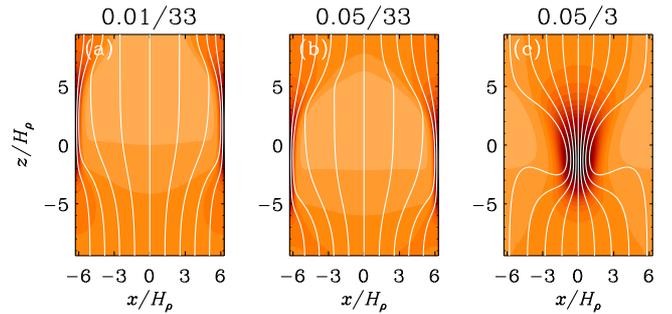}
\end{center}\caption[]{
Comparison of magnetic field structure in axisymmetric MFS.
(a) Run~0.01/33 with $B_0/\Beqz=0.01$ and
(b) Run~0.05/33 with $B_0/\Beqz=0.05$, both with $\kf H_\rho=33$.
The flow speeds vary from $-0.27\urms$ to $0.08\urms$ in both cases.
(c) Run~0.05/3 with $B_0/\Beqz=0.05$ and $\kf H_\rho=3$.
The flow speed varies from $-0.23\urms$ to $0.07\urms$.
}\label{ppfline_comp_wide}
\end{figure}

In MFS, as noted above,
the wavenumber of turbulent eddies, $\kf$, enters the expression
for the turbulent diffusivity via $\etat\approx\urms/3\kf$, and thus
$\tilde\etat\approx\Ma/3\tilde\kf$, so we have
\EQ
\tilde\kf\equiv\kf H_\rho=H_\rho^2/3\tau\etat
=\Ma/3\tilde\etat,
\EN
where $\tau=H_\rho/\urms$ is the turnover time per scale height.
When $\urms$ is kept unchanged, smaller scale separation implies
a decrease of $\tilde\kf$, i.e., the size of turbulent
eddies in the domain is increased.
Earlier work has indicated that the growth
rate of the instability for horizontal magnetic field
decreases with decreasing $\tilde\kf$ \citep{BKKR12}.
However, we do not know whether this also causes a change in
the spot diameter, which would be plausible, or a change in
the depth at which NEMPI occurs.
In our MFS we have chosen $\Ma=0.1$ and $\tilde\etat$ corresponds to
$\tilde\kf\approx33$.
For $\tilde\etat=5\times10^{-3}$ we have $\tilde\kf\approx7$,
which is about the smallest scale separation for which NEMPI is still
possible in this geometry; see \Fig{ppfline_comp_etat}.
Interestingly, as $\tilde\kf$ is decreased,
the location of the flux tube structure moves upward.
This can be understood as a consequence of enhanced turbulent
diffusion, which makes the flux tubes less concentrated, so the
magnetic field is weaker, but weaker magnetic field sinks less
than stronger fields.

Even for $\kf H_\rho\approx3$ it is still possible to find NEMPI in MFS,
but, as we have seen, the flux tube moves upward and becomes thicker.
To accommodate for this change, we need to increase the diameter of the
domain and, in addition, we would either need to extend it
in the upward direction or increase
the magnetic field strength to move the tube back down again; cf.\ \Fig{ppfline_comp}.
We choose here the latter.
In \Fig{ppfline_comp_wide}, we show three cases for a wider box.
In the first two runs (referred to as `0.01/33' and `0.05/33')
we keep the scale separation ratio the same as before,
i.e.\ $\tilde\kf=33$, and increase $B_0/\Beqz$ from 0.01 to 0.05,
while in the third case we keep $B_0/\Beqz=0.05$ and
decrease $\tilde\kf$ to 3.
We increase the magnetic field by a factor of 5 so as to keep
the structure within the computational domain.
In the first case, the natural separation between tubes would
be too small for this large cylindrically symmetric container.
By contrast, in a 3-D Cartesian domain, a second
downdraft would form, which is not possible in an axisymmetric geometry.
Instead, a downdraft develops on the outer rim of the container.
On the other hand, if $\tilde\kf$ is decreased and thus $\tilde\etat$ increased,
a single downdraft is again possible, as shown in \Fig{ppfline_comp_wide}(b),
suggesting that the horizontal scale of structures is also increased as
$\tilde\etat$ is increased.
We see that the tube can now attain significant diameters.
Its height remains unchanged, so the aspect ratio
of the structure is decreased as the scale separation ratio
is decreased.

\begin{figure}\begin{center}
\includegraphics[width=.94\columnwidth]{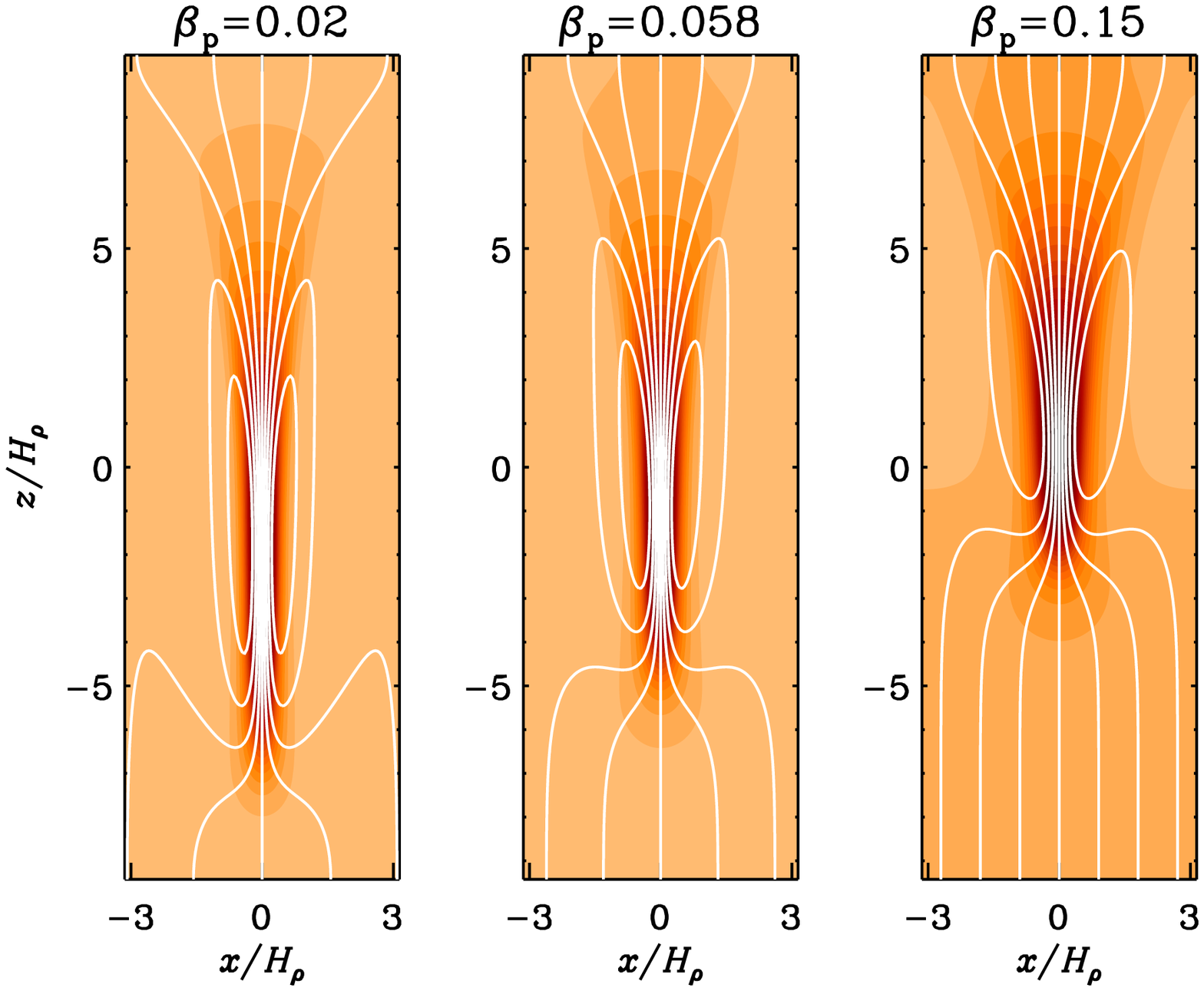}
\end{center}\caption[]{
Comparison of magnetic field structure in axisymmetric MFS
for Runs~Bu01/33--Bw01/33 with three values of $\betap$.
}\label{ppfline_comp_betap}
\end{figure}

\begin{figure}\begin{center}
\includegraphics[width=.94\columnwidth]{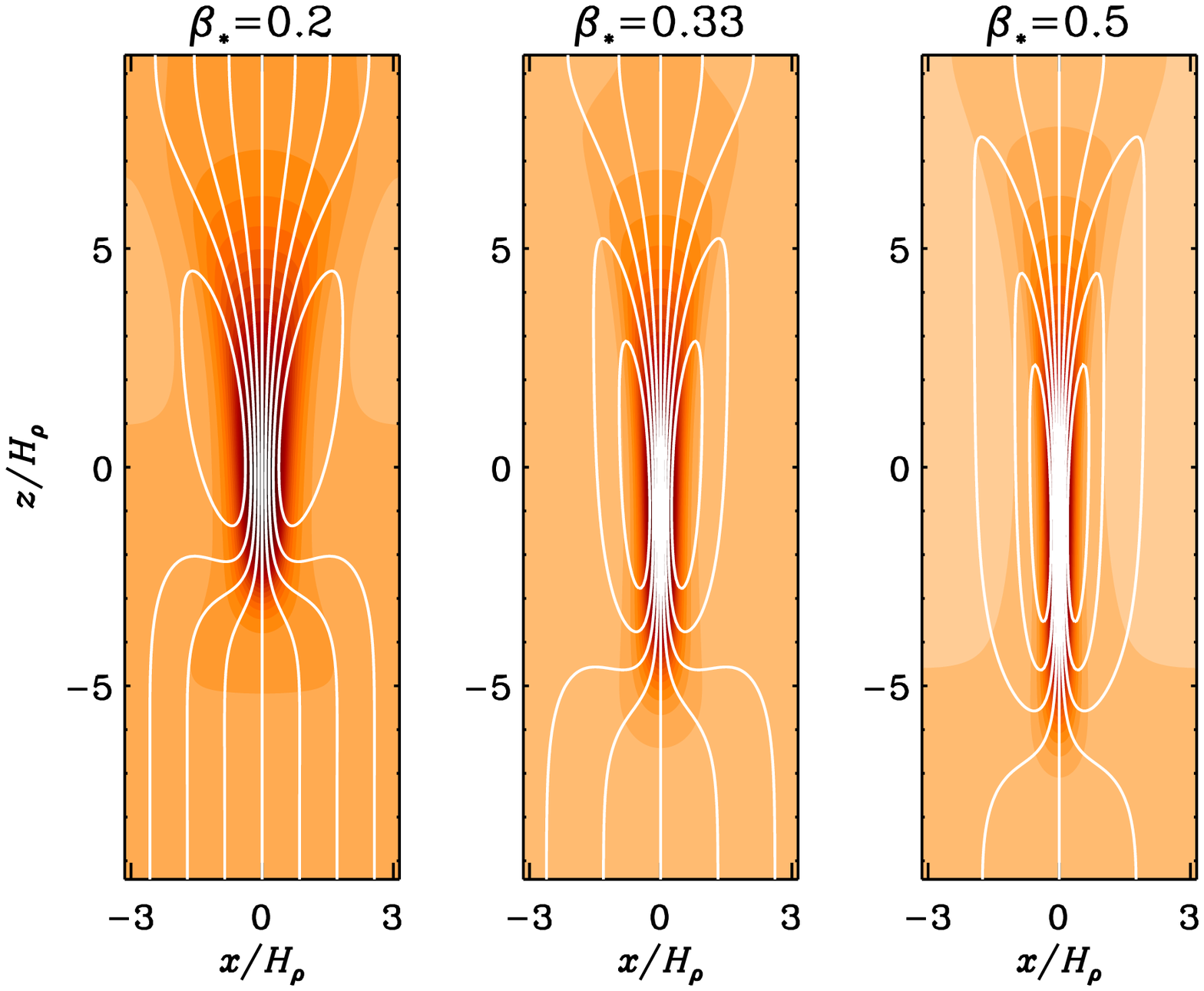}
\end{center}\caption[]{
Comparison of magnetic field structure in axisymmetric MFS
for Runs~Av01/33--Cv01/33 three values of $\betastar$.
}\label{ppfline_comp_betastar}
\end{figure}

\begin{table*}[t!]\caption{
Survey of axisymmetric MFS giving normalized growth rates, mean field strengths,
mean flow speeds, and other properties for different values of
$\beta_0$, $\betastar$, $\betap$, and $\tilde\kf$.
In all cases we have $k_1 H_\rho=1$, so $\tilde\kf=\hat\kf$.
Asterisks indicate that the domain was clipped at $z_{\rm bot}=0$.
}\vspace{12pt}\centerline{}\begin{tabular}{llcllllrcrcccrcccr}
Run & $\;\;\beta_0$ & $\!\qpz\!$ & $\betastar$ & $\betap$ &
$\betamin$ & $\;\;\Pmin$ & $\tilde\kf$ &
$\hat{\lambda}$ & $\!\hat{\meanB}_z^{\max}\!\!\!$ &
$\!\hat{\meanU}_z^{\min}\!\!\!$ & $\!\hat{\meanU}_z^{\max}\!$ &
$\tilde{z}_B$ & $\tilde{z}_B^{\rm NL}$ & $\tilde{Z}_{\rm b}$ &
$\tilde{Z}_{\rm t}$ & $\tilde{R}$ & $A$ \\
\hline
Ov002/33$\!\!$&0.002$\!\!$&32&0.33&0.058&0.125&$-0.036$&33 & 5.0 & 52 & $-0.27$ & 0.03 & 8.3 & $ 4.8$ & 3.5 & 4.6 &  0.27 & 17 \\ 
Ov01/33 & 0.01 & 32 & 0.33 & 0.058 & 0.125 & $-0.036$ & 33 & 5.6 & 52 & $-0.27$ & 0.08 & 5.0 & $ 1.6$ & 3.6 & 4.3 &  0.27 & 16 \\ 
Ov05/33 & 0.05 & 32 & 0.33 & 0.058 & 0.125 & $-0.036$ & 33 & 2.2 & 51 & $-0.27$ & 0.08 & 1.7 & $-1.7$ & 3.7 & 4.3 &  0.27 & 16 \\ 
\hline
Bv002/33$\!\!$&0.002$\!\!$&32&0.33&0.058&0.125&$-0.036$&33 & 7.6 & 52 & $-0.27$ & 0.03 & 7.9 & $ 2.0$ & 3.1 & 4.1 &  0.35 & 12 \\ 
Bv01/33 & 0.01 & 32 & 0.33 & 0.058 & 0.125 & $-0.036$ & 33 & 9.4 & 52 & $-0.27$ & 0.08 & 3.9 & $-1.2$ & 3.1 & 4.1 &  0.35 & 12 \\ 
Bv05/33 & 0.05 & 32 & 0.33 & 0.058 & 0.125 & $-0.036$ & 33 &12.3 & 51 & $-0.27$ & 0.08 & 1.8 & $-4.4$ & 3.0 & 4.1 &  0.35 & 12 \\ 
\hline
Bv01/33 & 0.01 & 32 & 0.33 & 0.058 & 0.125 & $-0.036$ & 33 & 9.4 & 52 & $-0.27$ & 0.08 & 4.8 & $-1.2$ & 3.1 & 4.1 &  0.35 & 12 \\ 
Bv01/17 & 0.01 & 32 & 0.33 & 0.058 & 0.125 & $-0.036$ & 17 & 2.1 & 25 & $-0.27$ & 0.08 & 4.9 & $ 0.3$ & 2.8 & 4.1 &  0.50 &  8 \\ 
Bv01/7  & 0.01 & 32 & 0.33 & 0.058 & 0.125 & $-0.036$ &  7 & 0.4 &  8 & $-0.23$ & 0.07 & 4.7 & $ 2.7$ & 2.4 & 3.8 &  0.95 &  4 \\ 
\hline
Bu01/33&0.01&$\!$270$\!$&0.33&0.02 & 0.079 & $-0.048$ & 33 & 5.1 & 69 & $-0.35$ & 0.10 & 4.5 & $-2.1$ & 4.0 & 4.5 &  0.30 & 15 \\ 
Bv01/33 & 0.01 & 32 & 0.33 & 0.058 & 0.125 & $-0.036$ & 33 & 9.4 & 52 & $-0.27$ & 0.08 & 4.8 & $-1.2$ & 3.1 & 4.1 &  0.35 & 12 \\ 
Bw01/33 & 0.01 & 4.8& 0.33 & 0.15  & 0.164 & $-0.016$ & 33 & 3.2 & 25 & $-0.15$ & 0.05 & 5.4 & $ 0.3$ & 2.2 & 3.8 &  0.50 &  8 \\ 
\hline
Av01/33 & 0.01 & 12 & 0.2  & 0.058 & 0.091 & $-0.010$ & 33 & 2.6 & 22 & $-0.13$ & 0.04 & 4.3 & $-0.3$ & 2.3 & 3.8 &  0.55 &  7 \\ 
Bv01/33 & 0.01 & 32 & 0.33 & 0.058 & 0.125 & $-0.036$ & 33 & 9.4 & 52 & $-0.27$ & 0.08 & 4.8 & $-1.2$ & 3.1 & 4.1 &  0.35 & 12 \\ 
Cv01/33 & 0.01 & 74 & 0.5  & 0.058 & 0.160 & $-0.097$ & 33 &10.6 & 91 & $-0.47$ & 0.09 & 5.4 & $-1.7$ & 3.5 & 4.3 &  0.25 & 17 \\ 
\hline
Av01/33*& 0.01 & 12 & 0.2  & 0.058 & 0.091 & $-0.010$ & 33 & 2.4 & 11 & $-0.07$ & 0.04 & 4.2 & $ 1.6$ & 1.4 & 3.4 &  0.85 &  4 \\ 
Bv01/33*& 0.01 & 32 & 0.33 & 0.058 & 0.125 & $-0.036$ & 33 & 4.8 & 21 & $-0.15$ & 0.08 & 4.7 & $ 1.5$ & 1.3 & 3.3 &  0.60 &  5 \\ 
Cv01/33*& 0.01 & 74 & 0.5  & 0.058 & 0.160 & $-0.097$ & 33 & 8.7 & 33 & $-0.26$ & 0.10 & 5.3 & $ 1.4$ & 1.2 & 3.2 &  0.50 &  6 \\ 
\label{Summary_betap_etc}\end{tabular}
\end{table*}

\subsection{Parameter sensitivity}

It is important to know the dependence of the solutions
on changes of the parameters $\betap$ and $\betastar$ that determine
the function $\qp$.
In \Figs{ppfline_comp_betap}{ppfline_comp_betastar},
we present results
where we change either $\betap$ or $\betastar$, respectively.
Characteristic properties of these solutions are summarized in
\Tab{Summary_betap_etc}.
Runs~Ov002/33--Ov05/33 are 2-D Cartesian
while all other ones are 2-D axisymmetric.
In addition to $\betap$ and $\betastar$, we also list the values of
$\qpz=\betastar^2/\betap^2$, as well as the minimum position of the
$\Peff(\beta)$ curve, namely \citep[cf.][]{KBKR12}
\EQ
\Pmin=-\half(\betastar^2-\betap^2)^2,\quad
\betamin=\left(\betap\sqrt{-2\Pmin}\right)^{1/2}.
\EN
The main output parameters include the normalized growth rate in the
linear regime, $\hat\lambda=\lambda H_\rho^2/\etat$,
the maximum normalized vertical field
in the tube
\EQ
\hat{\meanB}_z^{\max}=\left.\meanB_z^{\max}\right/B_0,
\EN
the minimum and maximum normalized velocities,
\EQ
\hat{\meanU}_z^{\min}=\left.\meanU_z^{\min}\right/\urms,\quad\quad
\hat{\meanU}_z^{\max}=\left.\meanU_z^{\max}\right/\urms,
\EN
the normalized maximum magnetic field positions in the linear and
nonlinear regimes, $\tilde{z}_B=z_B/H_\rho$ and
$\tilde{z}_B^{\rm NL}=z_B^{\rm NL}/H_\rho$, respectively,
the similarly normalized positions where $\meanB_z$ has dropped by
$1/e$ of its maximum at the bottom end $\tilde{Z}_{\rm b}$,
at the top end $\tilde{Z}_{\rm t}$, and to the side $\tilde{R}$ of the tube,
as well as the aspect ratio $A=Z_{\rm t}/R$.

\begin{figure}\begin{center}
\includegraphics[width=.90\columnwidth]{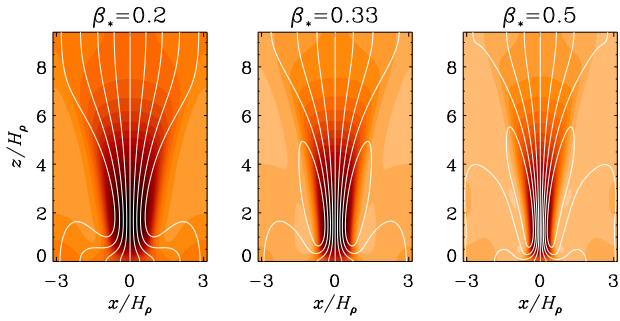}
\end{center}\caption[]{
Comparison of magnetic field structure in axisymmetric MFS
for Runs~Av--Cv01/33* with three values of $\betastar$
in a domain that is truncated from below.
}\label{ppfline_comp_betastar_3pi}
\end{figure}

The changes of $\hat\lambda$ are often as expected:
a decrease with decreasing values of $\tilde\kf$, and
a increase with increasing values of $\betastar$.
There are also some unexpected changes that could be associated
with the tube not being fully contained within our fixed domain:
for Run~Ov05/33 the domain may not be deep enough and
for Run~Bw01/33 it may not be wide enough.
Furthermore, we find that structures become taller when $\betap$ is small and
$\betastar$ large, and they become shorter and fatter when
$\betap$ is large and $\betastar$ small.
Thus, thicker structures, as indicated by the DNS of \cite{BKR13},
could also be caused by larger values of $\betap$ or smaller values
of $\betastar$.
When the domain is clipped at $z=0$, flux concentrations cannot fully
develop.
The structures are fatter and less strong; see \Fig{ppfline_comp_betastar_3pi}.

\subsection{Effect of rotation}

\begin{figure}\begin{center}
\includegraphics[width=.90\columnwidth]{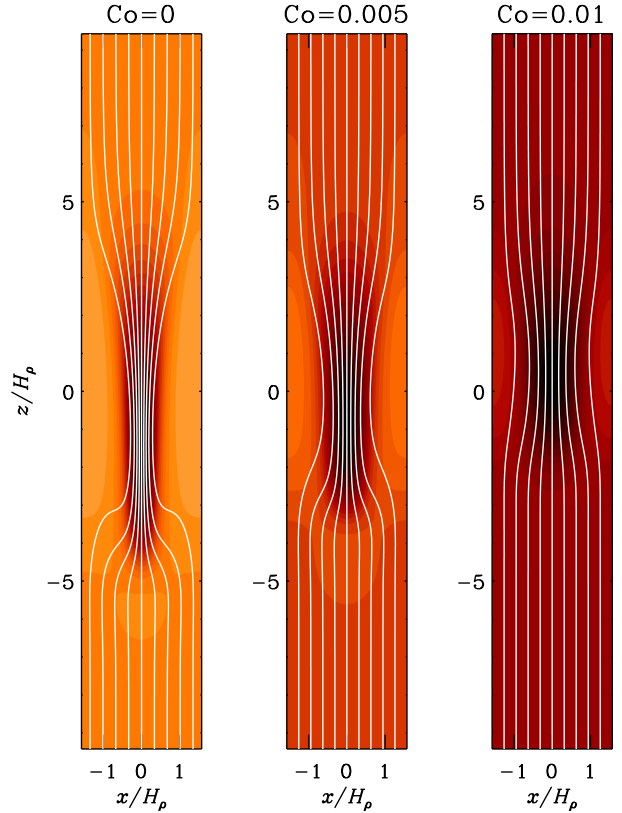}
\end{center}\caption[]{
Comparison of magnetic field structure in axisymmetric MFS
for a run similar to Run~Bv01/33, but for three values of $\Co$
and $r/H_\rho\le\pi/2$.
}\label{ppfline_comp_Om}
\end{figure}

The effect of rotation through the Coriolis force is determined
by the Coriolis number,
\EQ
\Co=2\Omega/\urms\kf=6\Omega\etat/\urms^2,
\EN
where $\Omega$ is the angular velocity.
\cite{LBKMR12,LBKR13} found that NEMPI begins to be suppressed
when $\Co\ga0.03$, which is a surprisingly small value.
They only considered the case of a horizontal magnetic field.
In the present case of a vertical magnetic field, we can use the
axisymmetric model to include a vertical rotation vector $\OO=(0,0,\Omega)$.
We add the Coriolis force to the right-hand side of \Eq{dUmean}, i.e.,
\EQ
\meanrho{\DD\meanUU\over\DD t}=...-2\OO\times\meanrho\meanUU.
\EN
When adding weak rotation ($\Co=0.01$) in Run~Bv01/33, it turns out
that magnetic flux concentrations develop on the periphery of the
domain, similar to the case considered in \Fig{ppfline_comp_wide}.
We have therefore reduced the radial extent of the domain to
$r/H_\rho\le\pi/2$.
The results are shown in \Fig{ppfline_comp_Om}.

In agreement with earlier studies, we find that rather weak
rotation suppresses NEMPI.
The magnetic structures become fatter and occur slightly higher
up in the domain.
For $\Co=0.01$, the magnetic flux concentrations have become rather weak.
If we write $\Co$ in terms of correlation of turnover time $\tau$
as $2\Omega\tau$, we find that the solar values of
$\Omega=3\times10^{-6}\s^{-1}$ corresponds to $30\min$.
According to stellar mixing length theory, this, in turn,
corresponds to a depth of less than 2\,Mm.

\section{DNS and ILES studies}

In the MFS discussed above, we have ignored the possibility of other terms
in the parameterization of the mean-field Lorentz force.
While this seems to capture the essence of earlier DNS \citep{BKR13},
this parameterization might not be accurate or sufficient in all respects.
It is therefore useful to perform DNS to see how the results depend on
scale separation, gravitational stratification, and Mach number.

\subsection{DNS and ILES models}
\label{Models}

We have performed direct numerical simulations using both the
{\sc Pencil Code} (\url{http://pencil-code.googlecode.com})
and {\sc Nirvana} (\url{http://nirvana-code.aip.de/}).
Both codes are fully compressible and are here used with an isothermal equation
of state with $p=\rho\cs^2$, where $\cs=\const$ is the sound speed.
The background stratification is then also isothermal.
Turbulence is driven using volume forcing given by a function $\ff$
that is $\delta$-correlated in time and monochromatic in space.
It consists of random non-polarized waves whose direction and
phase change randomly at each time step.

In DNS we solve the equations for the velocity $\UU$,
the magnetic vector potential $\AAA$, and the density $\rho$,
\begin{eqnarray}
{\DD\UU\over\DD t}&=&
-\cs^2\nab\ln\rho+{1\over\rho}\JJ\times\BB+\ff+\grav+\FF_\nu,
\\
{\partial\AAA\over\partial t}&=&\UU\times\BB+\eta\nabla^2\AAA,
\\
{\partial\rho\over\partial t}&=&-\nab\cdot\rho\UU,
\end{eqnarray}
where $\DD/\DD t=\partial/\partial t+\UU\cdot\nab$ is the advective
derivative, $\eta$ is the
magnetic diffusivity due to Spitzer conductivity of the plasma,
$\BB=\BB_0+\nab\times\AAA$ is the magnetic field,
$\BB_0=(0,0,B_0)$ is the imposed uniform vertical field,
$\JJ=\nab\times\BB/\mu_0$ is the current density,
$\mu_0$ is the vacuum permeability,
$\FF_\nu=\nab\cdot(2\nu\rho\SSSS)$ is the viscous force.
The turbulent rms velocity is approximately independent of $z$.
Boundary conditions are periodic in the horizontal
directions (so vertical magnetic flux is conserved),
and stress free on the upper and
lower boundaries, where the magnetic field is assumed to be
vertical, i.e., $B_x=B_y=0$.
In the ILES we solve the induction equation directly
for $\BB$, ignore the effects
of explicit viscosity and magnetic diffusivity and use an approximate
Riemann solver to keep the code stable and to dissipate kinetic
and magnetic energies at small scales.

The simulations are characterized by specifying a forcing amplitude,
which results in a certain rms velocity, $\urms$, and hence in a certain
Mach number.
Furthermore, the values of $\nu$ and $\eta$ are quantified through
the fluid and magnetic Reynolds numbers, $\Rey=\urms/\nu\kf$ and
$\Rm=\urms/\eta\kf$, respectively.
Their ratio is the magnetic Prandtl number, $\Pm=\nu/\eta$.
Occasionally, we also quote $\tilde\nu=\nu/\cs H_\rho$ and
$\tilde\eta=\eta/\cs H_\rho$.

An important diagnostics is the vertical magnetic field, $B_z$,
at some horizontal layer.
In particular, we use here the Fourier-filtered field,
$\meanB_z$, which is obtained by removing all components with wave
numbers larger than 1/6 of the forcing wavenumber $\kf$.
This corresponds to a position in the magnetic energy spectrum
where there is a local minimum, so we have some degree of
scale separation between the forcing scale and the scale of the spot.
We return to this in \Sec{DependenceScaleSeparation}.
To identify the magnetic field in the flux tube, we take
the maximum of $\meanB_z$, either at each height at one time,
which is referred to as $\meanB_z^{\max}(z)$,
or in the top layer at different times.
The latter is used to determine the growth rate of the instability.

When comparing results for different values of $g$,
it is convenient to keep the typical density at the surface the same.
Since our hydrostatic stratification is given by \Eq{meanrho_prof},
this is best done by letting the domain terminate at $z=0$ and to
consider the range $-L_z\leq z\leq0$.
In most of the cases we consider $L_z=\pi/k_1$,
although this might in hindsight be a bit short in some cases.
For comparison with earlier work of \cite{BKR13}, we also present models
in a domain $-\pi\leq k_1 z\leq\pi$.

\begin{table}[!t]\caption{
Summary of DNS at varying $\tilde{B}_0$, and fixed values of
$\tilde\eta=2\times10^{-4}$, $\Pm=0.5$, $\Rey\approx38$,
$\Ma\approx0.1$, $\hat{g}=1$, $\hat{k}_{\rm f}=30$,
$\tautd/\tauto\approx2700$, using $256^3$ mesh points.
In all cases the number of resulting spots is unity.
The positions $\tilde{z}_B^{\rm NL}$ agree with those marked in \Fig{pBprof_comp}.
}
\vspace{12pt}\centerline{}\begin{tabular}{ccccccccr}
$\!$Run$\!$ & $\hat{B}_0$ & $\Rey$ & $\Rm$ & $\Ma$ & $\hat{B}_z$ & $\hat{\meanB}_z$ &
$\tilde{R}$ & $\tilde{z}_B^{\rm NL}$ \\
\hline
(a) & 0.0005 & 39 &  19 & 0.12 &  1.81 & 0.36 & 0.13 & $ 3.1$ \\
(b) & 0.0010 & 39 &  19 & 0.12 &  2.68 & 1.00 & 0.11 & $ 1.8$ \\
(c) & 0.0020 & 38 &  19 & 0.11 &  2.45 & 0.87 & 0.17 & $ 1.4$ \\
(d) & 0.0050 & 37 &  18 & 0.11 &  3.47 & 1.25 & 0.22 & $-0.5$ \\
(e) & 0.0100 & 35 &  18 & 0.11 &  3.95 & 1.49 & 0.29 & $-1.2$ \\
(f) & 0.0200 & 31 &  16 & 0.09 &  4.21 & 1.26 & 0.45&$-\pi\;\;$\\
\label{Summary_B0}\end{tabular}
\end{table}

\begin{figure}\begin{center}
\includegraphics[width=\columnwidth]{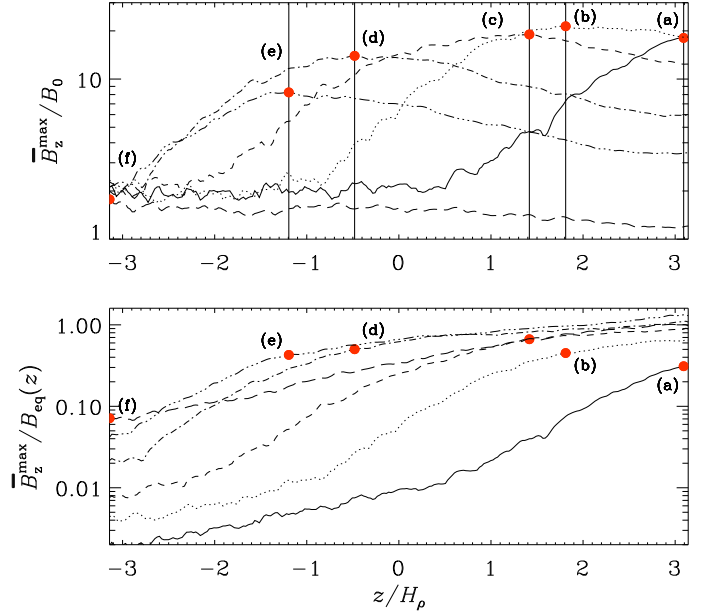}
\end{center}\caption[]{
Normalized vertical magnetic field profiles from DNS,
$\meanB_z^{\rm max}/B_0$ (top) and $\meanB_z^{\rm max}/\Beq(z)$ (bottom)
for the six values of $B_0/\Beqz$ listed in \Tab{Summary_B0}.
In both panels, the red dots mark the maxima of $\meanB_z^{\rm max}/B_0$
at positions $\tilde{z}_B^{\rm NL}$.
The labels (a)--(f) correspond to those in \Tab{Summary_B0}.
}\label{pBprof_comp}
\end{figure}

\subsection{Magnetic field dependence}

In \Tab{Summary_B0} and \Fig{pBprof_comp} we compare results
for six values of $\tilde{B}_0=B_0/(\mu_0\rho_0\csz^2)^{1/2}$.
These models are the same as those discussed in \cite{BKR13},
where visualizations are shown for all six cases.
Increasing $\tilde{B}_0$ leads to a decrease in the Mach number $\Ma$
and hence to a
mild decline of $\Rey$ and $\Rm$ for $\tilde{B}_0>0.01$, corresponding
to $B_0/\Beqz>0.1$.
There is a slight increase of $\tilde{B}_z^{\max}$,
while $\tilde{\meanB}_z^{\max}$ remains of the order of unity.
This is the case even for the largest value, $\tilde{B}_0=0.02$,
when NEMPI is completely suppressed and there is no distinct maximum
of $\meanB_z^{\max}/B_0$ in the upper panel of \Fig{pBprof_comp}.
This is why the visualization in \cite{BKR13} was featureless for
$\tilde{B}_0=0.02$, even though $\tilde{B}_z/\Beq(z)\approx1$ at
$z=z_{\rm top}$.
Moreover, while $\tilde{B}_z^{\max}$ shows only a slight increase,
the non-dimensional radius of the spot increases from
0.1 to about 0.4 as $\tilde{B}_0$ is increased.

\subsection{Magnetic Prandtl number dependence}
\label{MagneticPrandtl}

The results for different values of $\Pm$ are summarized in \Tab{Summary_Pm}.
It turns out that for $\Pm\ge5$,
no magnetic flux concentrations are produced.
We recall that analysis based on the quasi-linear approach (which is valid for
small fluid and magnetic Reynolds numbers) has shown that
for $\Pm\ge8$ and $\Rm \ll 1$,
no negative effective magnetic pressure is possible \citep{RKS12,BKKR12}.
Because of this, most of the earlier work used $\Pm=0.5$
so as to stay below unity in the hope that this would be
a good compromise between $\Pm$ being small and $\Rm$
still being reasonably large.
In fact, it now turns out that the difference in $\meanB_z^{\max}$
for $\Pm=1$ and 1/2 be negligible, and even for $\Pm=2$
the decline in $\meanB_z^{\max}$ is still small.
For $\Pm=0.2$, on the other hand, we find a large value of $\hat\lambda$,
but a low saturation level.
Again, this might be explained by the fact that the domain
is not deep enough in the $z$ direction, which can suppress NEMPI.
Alternatively, the resolution of $256^3$ might not be sufficient to resolve
the longer inertial range for smaller magnetic Prandtl numbers.
In \Sec{ReynoldsDependence} we present another case with $\Pm=0.2$
where both the resolution and the Reynolds numbers are doubled,
and $\meanB_z^{\max}$ is again large.

\begin{figure}[t!]\begin{center}
\includegraphics[width=\columnwidth]{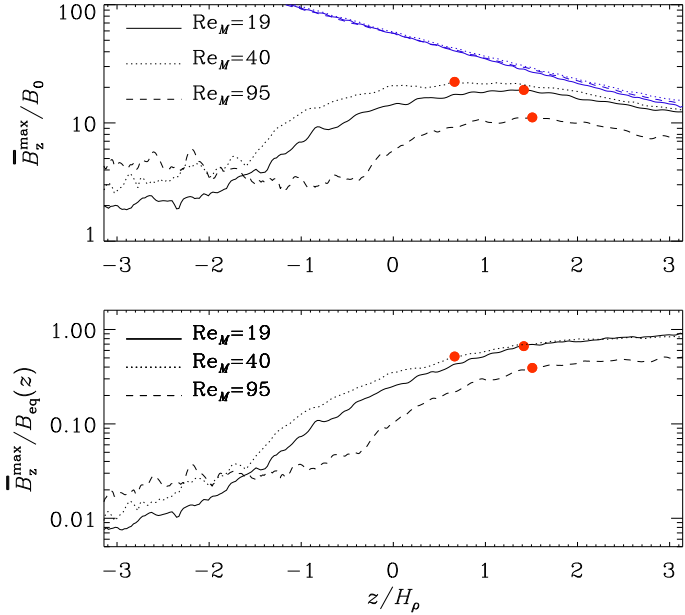}\\[4pt]
\end{center}\caption[]{
Similar to \Fig{pBprof_comp}, but for DNS Runs~A30/1--C30/1
listed in \Tab{Summary_Rm}, i.e., $\tilde\nu=10^{-4}$,
$\tilde{B}_0=0.002$, $\Pm=0.5$, $k_1 H_\rho=1$, and $\kf H_\rho=30$.
In the upper panel, the blue lines denote $\Beq(z)/B_0$ and in
both panels, the red dots mark the maxima of $\meanB_z^{\rm max}/B_0$
at positions $\tilde{z}_B^{\rm NL}$.
}\label{pBprof_comp_Rm}
\end{figure}

\begin{table}[!t]\caption{
Summary of DNS at varying $\Pm$, and fixed values of
$\tilde\eta=2\times10^{-4}$, $\tilde{B}_0=0.002$,
$\Rm\approx20$--$40$, $\Ma\approx0.1$, $\hat{g}=1$, $\hat{k}_{\rm f}=30$,
$\tautd/\tauto\approx2700$, using $256^3$ mesh points.
In all cases the number of spots is unity.
}\vspace{12pt}\centerline{}\begin{tabular}{cccccccc}
$\tilde{\nu}$ & $\!\!\Pm$ & $\Rm$ & $\Ma$ & $\hat{\lambda}$ & $\hat{B}_z$ &
$\hat{\meanB}_z$ & $\tilde{R}$ \\
\hline
$4\times10^{-5} $& 0.2 & 20 & 0.12 &  5.18 & 1.87 & 0.19 & 0.36 \\
$1\times10^{-4} $& 0.5 & 19 & 0.11 &  1.33 & 2.45 & 0.87 & 0.17 \\
$2\times10^{-4} $& 1   & 17 & 0.10 &  1.66 & 2.76 & 0.84 & 0.17 \\
$4\times10^{-4} $& 2   & 14 & 0.08 &  1.46 & 2.78 & 0.64 & 0.20 \\
$5\times10^{-4} $& 5   & 25 & 0.07 &  0.10 & 2.66 & 0.22 & 0.34 \\
$1\times10^{-3} $&10   & 19 & 0.06 &  0.04 & 2.87 & 0.28 & 0.30 \\
$5\times10^{-4} $&10   & 45 & 0.07 &  0.04 & 2.96 & 0.22 & 0.34 \\
\label{Summary_Pm}\end{tabular}
\end{table}

\begin{table}[!t]\caption{
Summary of DNS at varying $\Rey$ and $\Rm$, and fixed values of
$\tilde\nu=10^{-4}$, $\tilde{B}_0=0.002$,
$\Pm=0.5$, $k_1 H_\rho=1$, and $\kf H_\rho=30$.
}\vspace{12pt}\centerline{}\begin{tabular}{crcccccc}
Run & $\Rey$ & $\!\Rm\!$ & $\Ma$ & $\hat{B}_z$ & $\hat{\meanB}_z$ &
$\tilde{R}$ & Resol.\\
\hline
A30/1 & 38 & 19 & 0.11 & 2.45 & 0.87 & 0.17 & $256^3$\\
B30/1 & 80 & 40 & 0.12 & 3.30 & 1.02 & 0.16 & $512^3$\\
b30/1 &200 & 40 & 0.12 & 3.45 & 1.10 & 0.15 & $512^3$\\
C30/1 &190 & 95 & 0.11 & 3.47 & 0.71 & 0.19 & $1024^3$\\
D30/1 &190 & 95 & 0.11 & 3.54 & 0.69 & 0.19 & $\!\!1024^2\!\times\!1536\!\!$\\
E30/1 &190 &190 & 0.11 & 3.23 & 0.39 & 0.25 & $\!\!1024^2\!\times\!1536\!\!$\\
\label{Summary_Rm}\end{tabular}
\end{table}

\subsection{Reynolds number dependence}
\label{ReynoldsDependence}

\begin{figure*}[t!]\begin{center}
\includegraphics[width=\textwidth]{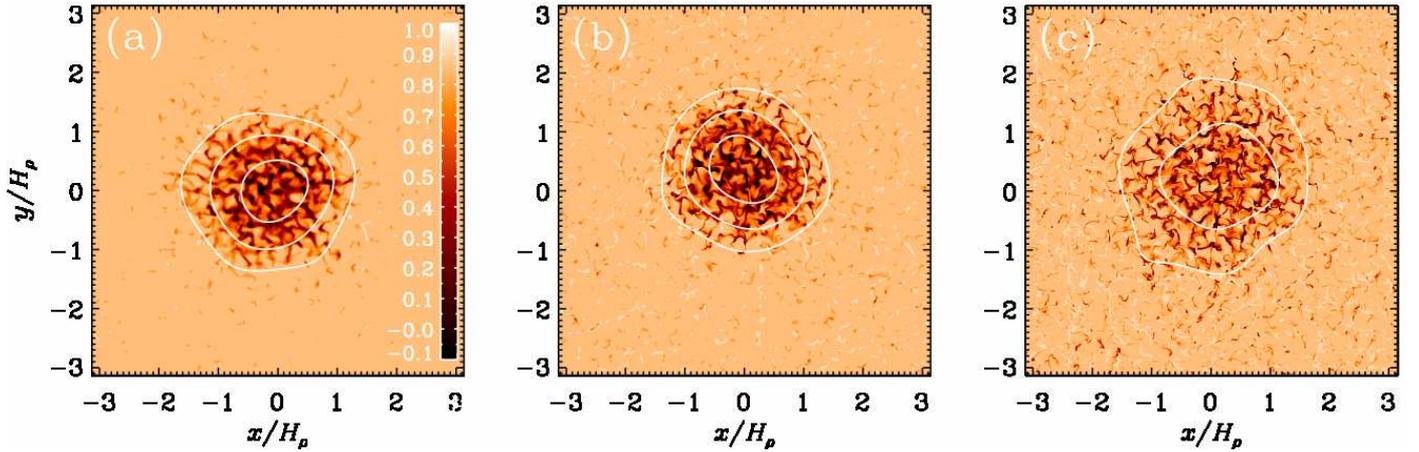}\\[4pt]
\end{center}\caption[]{
Magnetic field configuration at the upper surface
for DNS Runs~A30/1--C30/1 at three values of the magnetic Reynolds number.
The white contours represent the Fourier-filtered
with $k_\perp\leq\kf/6$; their levels correspond to
$\meanB_z^{\max}/\Beq(z_{\rm top})=0.05$, 0.2, and 0.4.
}\label{pBzm_top_comp}
\end{figure*}

Increasing $\Rm$ from 19 to 95, we see some changes; see \Tab{Summary_Rm}.
There is first a small increase of $\meanB_z^{\max}/\Beq(z)$ from 0.87 to 1.02
as $\Rm$ is increased from 19 to 40 (Run~B30/1).
Increasing $\Rey$ to 200, but keeping $\Rm=40$, results in a
further increase of $\meanB_z^{\max}/\Beq(z)$ to 1.10 (Run~b30/1).
This is also an example of a strong flux concentration with $\Pm=0.2$;
cf.\ \Sec{MagneticPrandtl}.
However, when $\Rm$ is increased further
to 95, $\meanB_z^{\max}/\Beq(z)$ decreases to about 0.71; see \Tab{Summary_Rm}.
Again, the weakening of the spot might be a consequence of the
domain not being deep enough.
Alternatively, it could be related to the occurrence of small-scale
dynamo action, which is indicated by the fact that in deeper layers
the small-scale magnetic field is enhanced in the run with the largest
value of $\Rm$; see \Fig{pBprof_comp_Rm}.
In Run~C30/1 with the largest value of $\Rm$, the spot is larger and
more fragmented, but it still remains in place and statistically steady;
see \Fig{pBzm_top_comp} and \url{http://www.nordita.org/~brandenb/movies/NEMPI/}
for corresponding animations.

\begin{table}[!t]\caption{
Summary of DNS at varying $\hat\kf=\kf/k_1$, $\tilde\kf=\kf H_\rho$,
$\hat{g}=g/\cs^2 k_1$, and fixed values of $\tilde{B}_0=0.02$,
$\tilde{\eta}_0=2\times10^{-4}$,
using resolutions of $256^3$ mesh points (for Run~a30/1),
$512^3$ mesh points (for Run~a30/4, a10/3, and a30/3),
as well as $1024^2\times384$ mesh points (for Runs~a40/1 and A40/1).
}
\vspace{12pt}\centerline{}\begin{tabular}{cccccccccrc}
Run & $\Pm$ & $\Rm$ & $\Ma$ & $\hat\kf$ & $\tilde\kf$ & $\hat{g}$ &
$\hat\lambda$& $\hat{B}_z$ & $\hat{\meanB}_z$ \\
\hline
a30/1 & 1.0 &  16 & 0.09 & 30 & 30 & 1 & 0.94 & 3.09 & 0.78 \\
a30/4 & 1.0 &  21 & 0.13 & 30 & 7.5 & 4 & 0.18 & 4.42 & 0.88 \\
a10/3 & 1.0 &  63 & 0.13 & 10 & 3.4 & 3 & ---  & 4.83 & 0.40 \\
a40/1 & 1.0 &  33 & 0.07 & 40 & 10  & 1 & 0.83 & 3.83 & 0.87 \\
A40/1 & 1.0 &  33 & 0.07 & 40 & 10  & 1 & 1.05 & 5.81 & 1.41 \\
a30/3 & 0.5 &  23 & 0.14 & 30 & 10  & 3 & 0.46 & 4.47 & 1.31 \\
\label{Summary_gkf}\end{tabular}
\end{table}

\begin{figure}[t!]\begin{center}
\includegraphics[width=\columnwidth]{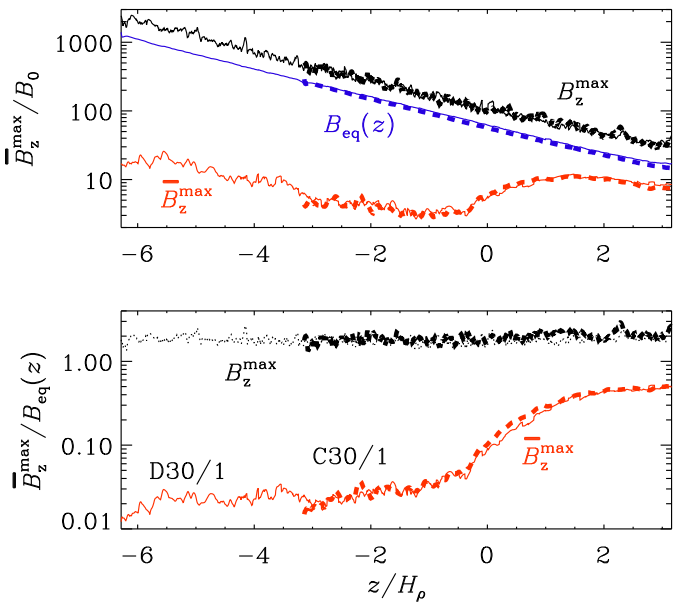}\\[4pt]
\end{center}\caption[]{
Similar to \Fig{pBprof_comp_Rm}, but for DNS Runs~C30/1
($z_{\rm bot}/H_\rho=-\pi$; thicker lines) and
D30/1 ($z_{\rm bot}/H_\rho=-1.5\pi$; thinner lines)
listed in \Tab{Summary_Rm}, i.e., $\tilde\nu=10^{-4}$,
$\tilde{B}_0=0.002$, $\Pm=0.5$, $k_1 H_\rho=1$, and $\kf H_\rho=30$.
In the upper panel, the blue lines denote $\Beq(z)/B_0$.
}\label{pbzmmax_comp}
\end{figure}

To eliminate the possibility of the domain not being deep enough,
we have performed additional simulations where we have extended the domain
in the negative $z$ direction down to $z_{\rm bot}/H_\rho=-1.5\pi$.
In \Fig{pbzmmax_comp} we show a computation of the resulting profiles of
$\meanB_z^{\max}/B_0$ and $\meanB_z^{\max}/\Beq(z)$.
We also include here a run with $\Pm=1$ instead of 0.5 (Run~E30/1).
It turns out that the strength of the spot is unaffected by the position
of $z_{\rm bot}$ and that there is a deep layer below $z/H_\rho\approx-2$
in which there is significant magnetic field generation owing to small-scale
dynamo action, preventing thereby also the value of $\meanB_z^{\max}/\Beq(z)$
to drop below the desired value of 0.01.
This might explain the weakening of the spot.
This is consistent with earlier analytical \citep{RK07} and
numerical \citep{BKKR12} work showing a finite drop of the important
NEMPI parameter $\betastar$ around $\Rm=60$.

\subsection{Dependence on scale separation and stratification}
\label{DependenceScaleSeparation}

\begin{figure*}[t!]\begin{center}
\includegraphics[width=\textwidth]{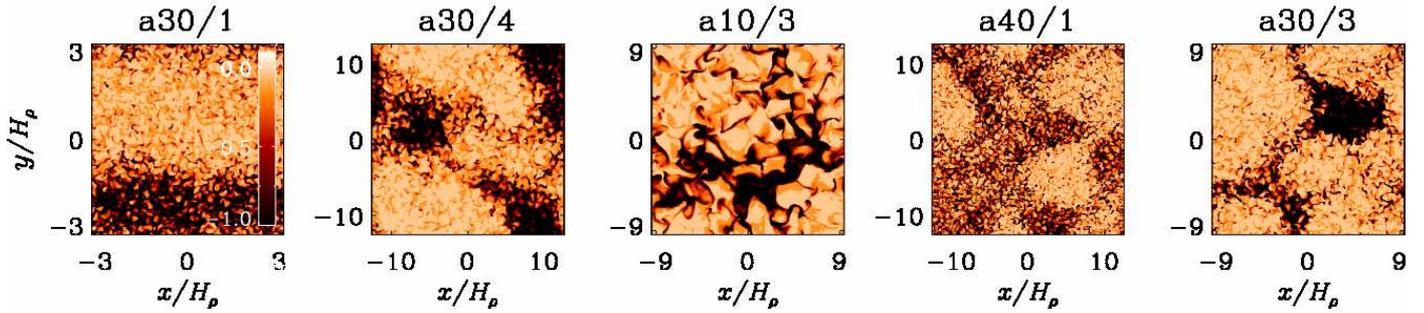}\\[4pt]
\end{center}\caption[]{
Magnetic field configuration at the upper surface
for DNS Runs~a30/1--a30/3 of \Tab{Summary_gkf}.
}\label{pBzm_top_comp_gkf}
\end{figure*}

We have performed various sets of additional simulations where we change
$\hat{g}$ and/or $\hat\kf$; see \Tab{Summary_gkf} and \Fig{pBzm_top_comp_gkf}.
In those cases, the vertical extent of the domain is from $-\pi$ to 0.
As discussed in \Sec{Models} this might be too small in some cases
for NEMPI to develop fully.
Nevertheless, in all cases there are clear indications
for the occurrence of flux concentrations.
The results regarding the growth rate of NEMPI
are not fully conclusive, because the changes in $\kf$ and
$H_\rho$ also affect turbulent--diffusive and turnover
time scales.
As shown in the appendix of \cite{KBKMR13}
the normalized growth rate of NEMPI is given by:
\EQ
\hat\lambda+1=3\betastar\,(\kf H_\rho)/(k_\perp H_\rho)^2,
\EN
which is not changed significantly for a vertical magnetic field;
see \Sec{GrowthRate}.
If $k_\perp H_\rho=\const\approx0.7$,
as suggested by the MFS of \Sec{AspectRatio}, we would expect
$\hat\lambda+1$ to be proportional to $\kf H_\rho$,
which is not in good agreement with the simulation results.

\begin{figure}[t!]\begin{center}
\includegraphics[width=\columnwidth]{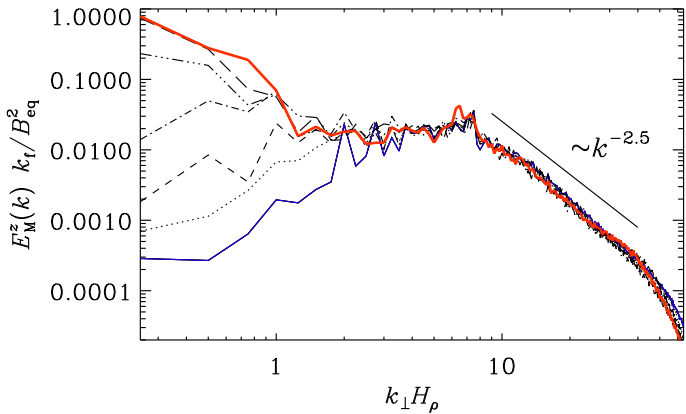}
\end{center}\caption[]{
Normalized spectra of $B_z$ from DNS Run~a30/4 at normalized times
$t\etatz/H_\rho^2\approx0.2$, 0.5, 1, 2, 5, 10, and 20,
for $\hat{g}\equiv g/\cs^2\kf=4$.
}\label{pBz_spec2}
\end{figure}

\begin{figure}[t!]\begin{center}
\includegraphics[width=\columnwidth]{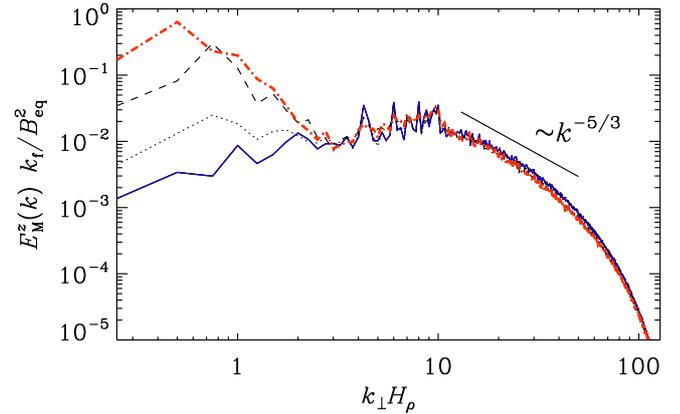}
\end{center}\caption[]{
Normalized spectra of $B_z$ from DNS Run~A40/1 at normalized times
$t\etatz/H_\rho^2\approx0.2$, 0.5, 1, and 2.7
with $\kf H_\rho=10$ and $k_1 H_\rho=0.25$.
}\label{pBz_spec2_halfV1024x384k10VF_Bz02_g1_pm1}
\end{figure}

To shed some light on this, we now discuss horizontal power
spectra of $B_z(x,y)$ taken at the top of the domain.
These spectra are referred to as $E_{\rm M}^z(k)$ and are
normalized by $\Beq^2/\kf$.
In Run~a30/4 with $\hat{g}=4$ and $\hat{k}_{\rm f}=30$,
we have access to wavenumbers down to $k_1 H_\rho=0.25$.
The results in \Fig{pBz_spec2} show that there is significant power
below $k_\perp H_\rho=0.7$.
This is in agreement with the MFS in the nonlinear regime;
see \Fig{pBz_spec2_256x256x128_16pi_B01a}.
The time evolution of $E_{\rm M}^z(k)$ suggest a behavior
similar to that of an inverse magnetic helicity cascade
that was originally predicted by \citep{FPLM75} and later
verified both in closure calculations \citep{PFL76} and DNS \citep{B01}.
Similar results with inverse spectral transfer
are shown in \Fig{pBz_spec2_halfV1024x384k10VF_Bz02_g1_pm1}
for DNS Run~A40/1.
The only difference between Runs~A40/1 and a40/1 is the vertical
extent of the domain, which is twice as tall in the former case
($3\pi H_\rho$ instead of $1.5\pi H_\rho$).
We note in this connection that the spectra tend to show a local
minimum near $\kf/6$.
This justifies our earlier assumption of separating mean and fluctuating
fields at the wavenumber $\kf/6$; see \Sec{Models}.
The spectra also show something like an inertial subrange proportional
to $k^{-5/3}$ (\Fig{pBz_spec2_halfV1024x384k10VF_Bz02_g1_pm1})
or $k^{-2.5}$ (\Fig{pBz_spec2}).
The latter is close to the $k^{-3}$ subrange in the MFS
of \Fig{pBz_spec2_256x256x128_16pi_B01a}.
Those steeper spectra could be a symptom of a low Reynolds number or,
alternatively, a consequence of most of the energy inversely `cascading'
to larger scales in the latter two cases.

We have also checked how different kinds of helicities
vary during NEMPI.
In the present case, magnetic and kinetic helicities are
fluctuating around zero, but cross helicity is not.
The latter is an ideal invariant of the magnetohydrodynamic (MHD) equations,
but in the present case of a stratified layer with a
vertical net magnetic field, $\bra{\uu\cdot\bb}$ can
actually be produced; see \cite{RKB11}, who showed that
\EQ
\bra{\uu\cdot\bb}\approx\etat\grav\cdot\BB_0/\cs^2
=-\etat B_0/H_\rho.
\EN
In a particular case of Run~30/1, we find a time-averaged value of
$\bra{\uu\cdot\bb}$ that would suggest that $\etat/\etatz$
is around 6, which is significantly larger than unity.
This would agree with independent arguments in favor of having
underestimated $\etat$; see the discussion in \Sec{VerticalProfile}.
In other words, if $\etat$ were really larger than what is
estimated based on the actual rms velocity, it would also
explain why the diameter of tubes is bigger in the DNS than in the MFS.

\begin{table}[!b]\caption{
Summary of DNS and ILES at varying values of $\Ma$,
all for $\hat{g}=3$, $\hat\kf=30$.
For ILES, no accurate values of $\hat\lambda$ are available.
In the DNS, the resolution is $256^2$ for Runs~D01 and D02,
and $512^2$ for Run~D10, while for Runs~I03--I30 it is $256^2\times128$.
}
\vspace{12pt}\centerline{}\begin{tabular}{ccccccc}
Run & $\tilde{B}_0$ & $\Rm$ & $\Ma$ & $\hat\lambda$& $\hat{B}_z$ & $\hat{\meanB}_z$ \\
\hline
D01 & 0.01 & 24 & 0.15 & 0.28 & 3.06 & 0.78 \\
D02 & 0.02 & 24 & 0.14 & 0.46 & 4.47 & 1.31 \\
D10 & 0.10 &  8 & 0.50 & 0.25 & 4.91 & 1.61 \\
I03 & 0.10 & ---& 0.16 & $>1$ & 2.86 & 1.14 \\
I10 & 0.10 & ---& 0.34 & $>1$ & 2.70 & 1.00 \\
I30 & 0.10 & ---& 0.68 & $>1$ & 2.41 & 1.02 \\
\label{Summary_Ma}\end{tabular}
\end{table}

\begin{figure}\begin{center}
\includegraphics[width=\columnwidth]{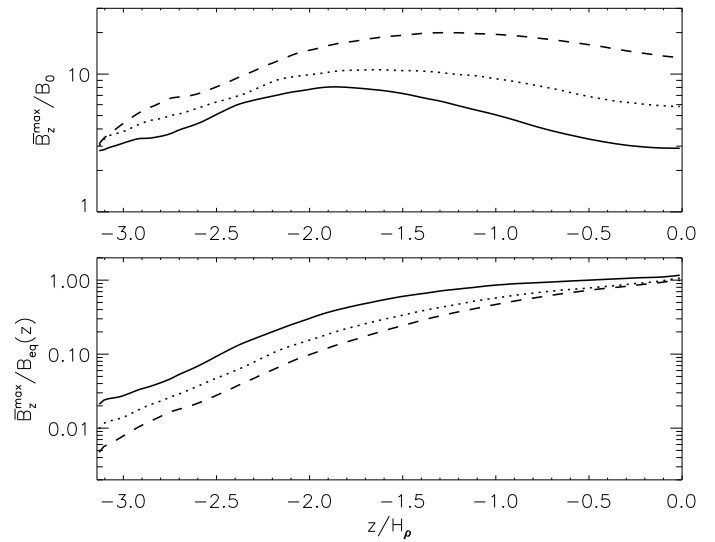}
\end{center}\caption[]{
Same as \Fig{pBprof_comp}, but for the ILES runs with varying forcing
amplitude. Note the different vertical extent of this set of
models. The different lines indicate $\Ma=0.16$ (solid), $0.34$
(dotted), and $0.68$ (dashed).}
\label{fig:pBprof_comp_iles}
\end{figure}

\begin{figure}\begin{center}
\includegraphics[width=\columnwidth]{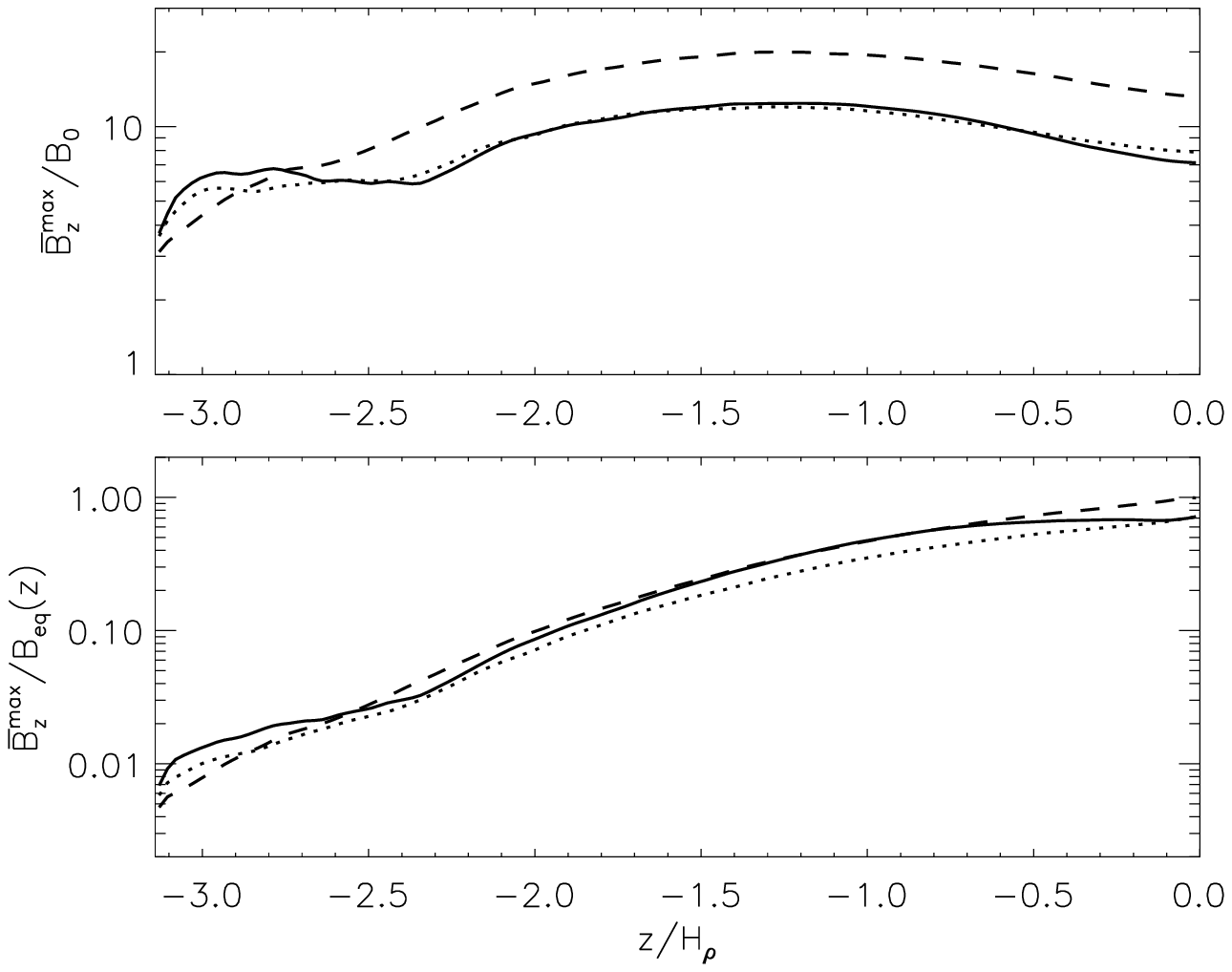}
\end{center}\caption[]{
Same as \Fig{fig:pBprof_comp_iles}, but for the ILES runs which vary
both the forcing amplitude and the imposed magnetic field at the same
time, keeping the \emph{relative} field strength comparable. The
different lines again indicate $\Ma=0.16$ (solid), $0.35$ (dotted),
and $0.68$ (dashed), and agree markedly well in the lower panel where
we plot relative to $\Beq(z)$.}
\label{fig:pBprof_comp_iles_alt}
\end{figure}

\subsection{Mach number dependence}

To study the dependence on Mach number, it is useful to consider ILES
without any explicit viscosity or magnetic diffusivity.  In
\Figss{fig:pBprof_comp_iles}{fig:pBzmax_Ma} we show the results for
three values of Ma at $\hat{g}=g k_1/\cs^2=3$ and $\hat\kf=\kf/k_1$
using a resolution of $256^2\times128$ nodes on the mesh.
In \Tab{Summary_Ma} we give a summary of various output parameters
and compare with corresponding DNS.
Note first of all that the
results from ILES are generally in good agreement with the DNS.  This
demonstrates that the mechanism causing magnetic flux concentrations
by NEMPI is robust and not sensitive to details of the magnetic
Reynolds number, provided that $\Rm \ga 10$.
The normalized growth rate is in all three cases above unity.

\begin{figure}[t!]\begin{center}
\includegraphics[width=0.48\columnwidth]{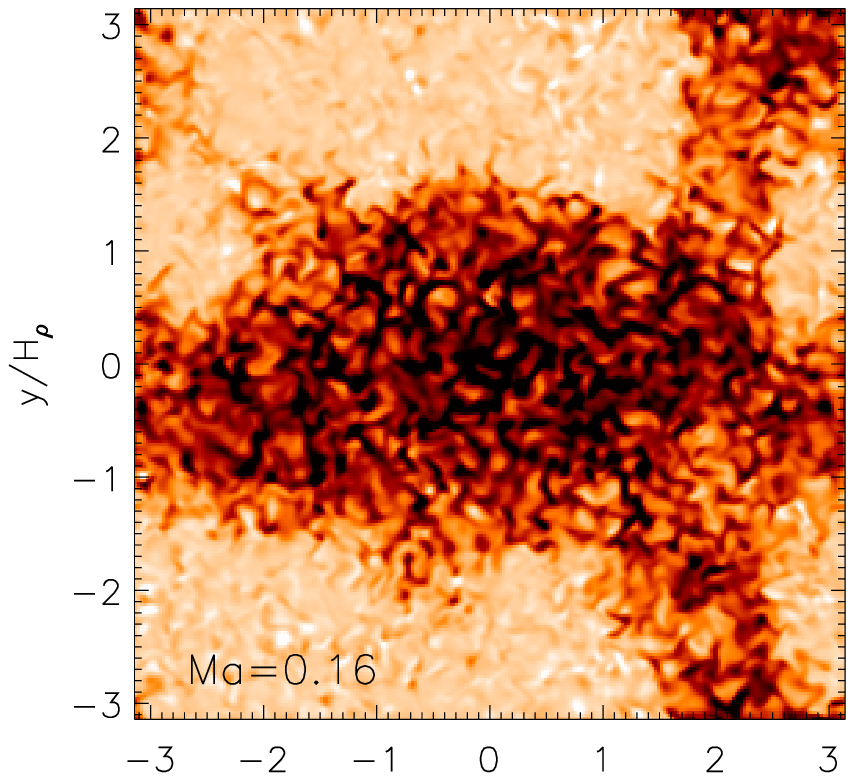}\hfill
\includegraphics[width=0.48\columnwidth]{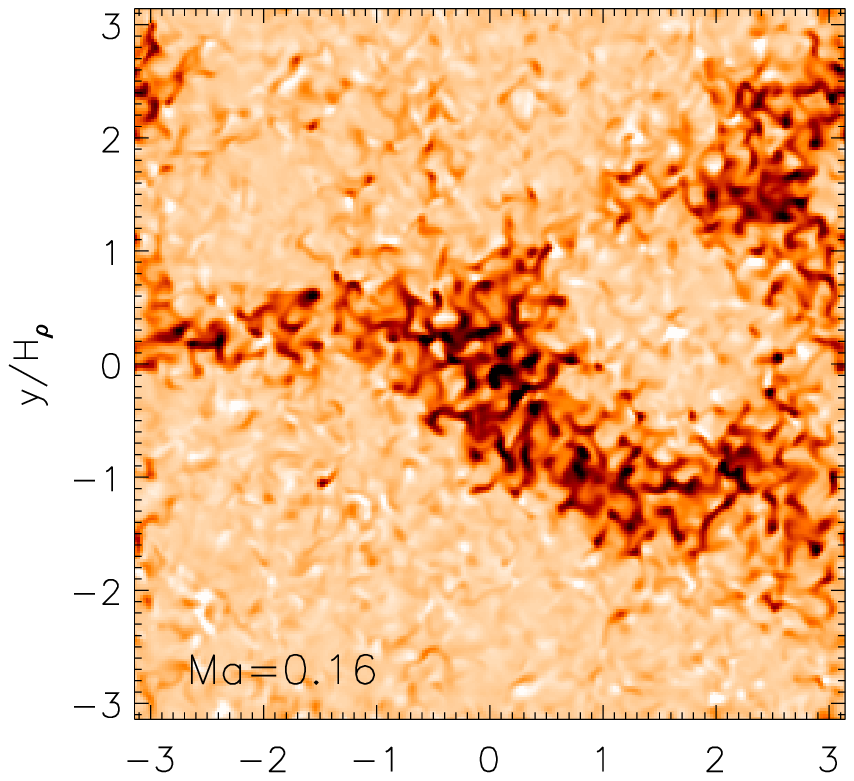}\\[4pt]
\includegraphics[width=0.48\columnwidth]{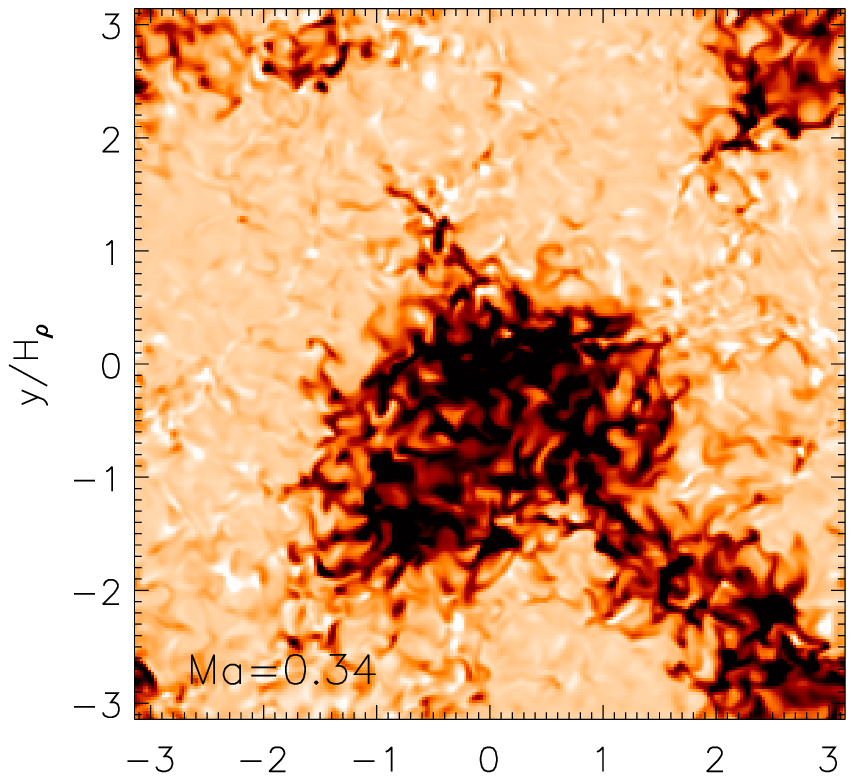}\hfill
\includegraphics[width=0.48\columnwidth]{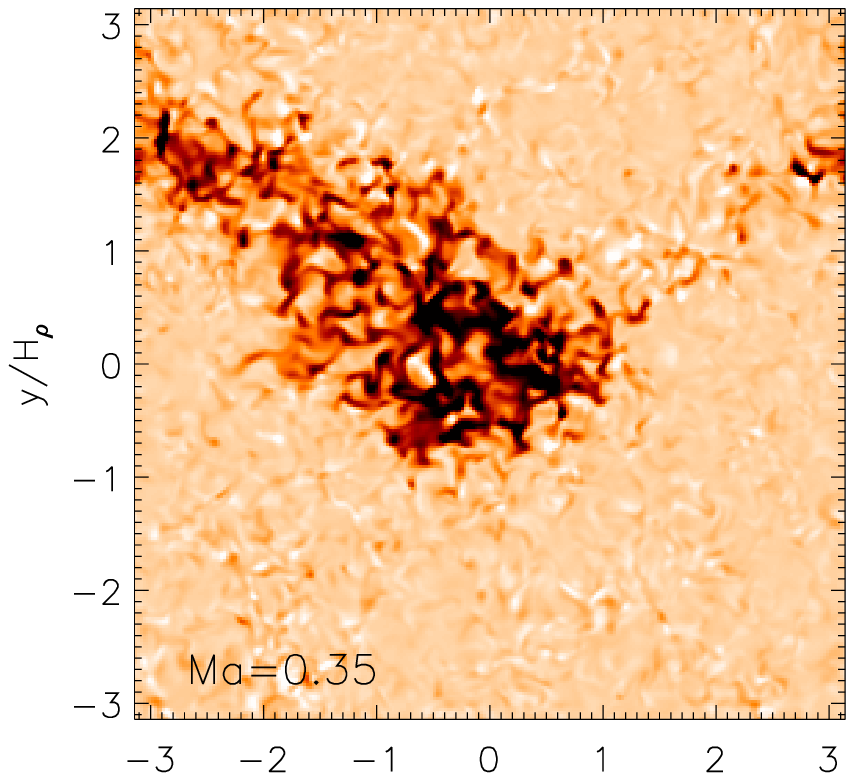}\\[4pt]
\includegraphics[width=0.48\columnwidth]{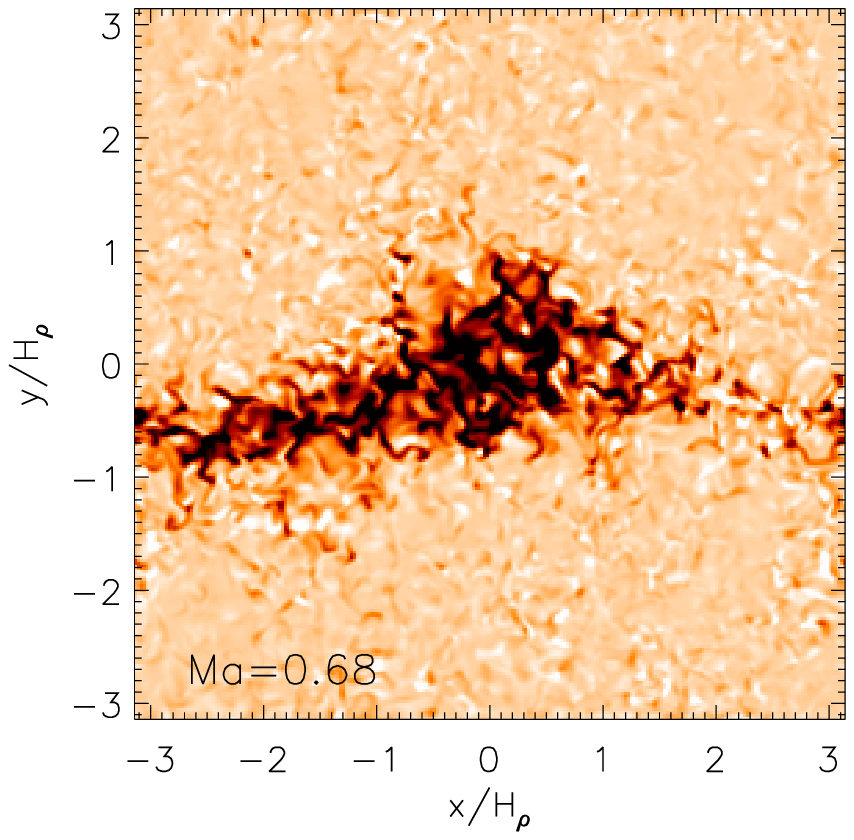}
\end{center}\caption[]{Surface appearance of the vertical magnetic
    field, $B_z^{\max}$, in the ILES simulations with different Mach
    numbers (top to bottom). The color coding shows $B_z^{\max}/\Beq$
    in the range of $-0.1$ (white) to $+1.0$ (black). Root-mean-square
    Mach numbers are given by the labels. For the upper two rows with
    lower Mach number, the left column is for fixed initial mean
    field, whereas in the right column the initial field is adjusted
    between the runs, such that the field strength remains constant
    relative to the kinetic energy in the background
    turbulence.}\label{fig:pBzmax_Ma}
\end{figure}

As the Mach number is increased, the magnetic structures become
smaller; seen in the left-hand panels of \Fig{fig:pBzmax_Ma}.  Since
the properties of NEMPI depend critically on the ratio $\meanB_z/\Beq$,
and since $\Beq(z)$ and hence $\Beqz$ increase with increasing Mach
number, the decrease in the size of magnetic structures might just
reflect the fact that for smaller values of $\meanB_z/\Beqz$, NEMPI
would operate at higher layers which are no longer included in our
computational domain.
Looking at \Fig{fig:pBprof_comp_iles}, it is clear that the maximum
of $\meanB_z/B_0$ moves to higher layers, but it is still well confined
within the computational domain.
Nevertheless, if one compensates for the decrease of $\Beq$ by applying
successively weaker mean fields when going to lower Mach number
(right-hand panels of \Fig{fig:pBzmax_Ma}), the size of the emerging
structures remains approximately similar and the curves of
$\meanB_z/\Beq$ lie now nearly on top of each other.
This shows clearly that in our
simulations with $\Ma\approx0.7$, flux concentrations are well
possible.  This is important, because it allows for the possibility
that the energy density of magnetic flux concentrations can become
comparable with the thermal energy.

\section{Possible application to sunspot formation}

Compared with earlier studies of NEMPI using a horizontal imposed
magnetic field, the prospects of applying it to the Sun have improved
in the sense that the flux concentrations are now stronger
when there is a vertical magnetic field.
In particular, the resulting magnetic structures survive in the
presence of larger Mach numbers up to 0.7, which is relevant to the
photospheric layers of the Sun \citep{SN98}.
However, those structures do become somewhat weaker
as the magnetic Reynolds number is increased sufficiently to allow
for the presence of small-scale dynamo action.
This was expected based on a certain drop of $\betastar$ for $\Rm\ga60$
found in earlier simulations \citep{BKKR12}, but the possibility of
spot formation still persists.
More specifically, we have seen that the largest field strength,
$\meanB_z/B_0$, occurs at a height where $\meanB_z^{\max}/\Beq$
is at least 0.4; see \Fig{pbzmmax_comp}.

To speculate further regarding the applicability to sunspot formation,
we must look at the mean-field models presented in \Sec{MFS}.
In particular, we have seen that spot formation occurs at a depth
$z_B^{\rm NL}$ where $\meanB_z^{\max}/\Beq$ is between 0.6 (for $\Rm=40$)
and 0.4 (for $\Rm=95$); see \Fig{pBprof_comp_Rm} and \Tab{Summary_Rm}.
Larger ratios of $\meanB_z^{\max}/\Beq$ occur in the upper layers,
but then the absolute field strength is lower.
In the MFS of \Sec{MFS}, the value of $\meanB_z^{\max}/\Beq$ at
$z=z_B^{\rm NL}$ is somewhat smaller (around 0.3), suggesting that the
adopted set of mean-field parameters in \Eq{qpz} was slightly suboptimal.
Nevertheless, those models show that the depth where NEMPI occurs and
where the effective magnetic pressure is most negative is even further
down, e.g., at $z/H_\rho\approx-7$; see \Fig{ppfline_flow}.
Furthermore, at the depth where $(-\dd\Peff/\dd\ln\beta^2)^{1/2}$
is maximum, i.e., where NEMPI is strongest according to theory,
we find $\meanB_z^{\max}/\Beq\approx0.05$.
Thus, there is an almost tenfold increase of the absolute field strength
between the depth were NEMPI occurs and where the field is strongest.

As we have seen from \Fig{ppfline_flow}, this increase is caused solely
by hydraulic effects, similar to what \cite{Par76,Par78} anticipated
over 35 years ago.
Our isothermal models clearly do demonstrate the hydraulic effect due
to downward suction, but we cannot expect realistic estimates for
the resulting field amplification.
\cite{Par78} gives more realistic estimates, but in his work the source
of downward flows remained unclear.
Our present work suggests that NEMPI might drive such motion,
but in realistic simulations it would be harder to identify this
as the sole mechanism.
Another mechanism might simply be large-scale hydrodynamic convection flows
that would continue deeper down to the lower part of the supergranulation
layer at depths between 20 and $40\,\Mm$.
Some indications of this have now been seen in simulations of \cite{SN12}.
Whether the reason for flux concentrations is then NEMPI or convection
can only be determined through careful numerical experiments comparing
full MHD with the case of a passive vector field.
Such a field would still be advected by
convective flows but would not contribute to the dynamical effects
that would be required if NEMPI were to be responsible.

In addition to the magnetic field strength of flux concentrations, there
might also be issues concerning their size.
Usually they are not much larger than about 5 density scale heights; see,
e.g., \Fig{pBzm_top_comp}.
This might be too small to explain sunspots.
On the other hand, in the supergranulation layer, the density
scale height increases, and larger scale structures might be produced
at those depths.

To make this more concrete, let us discuss a possible scenario.
At a depth of $3\Mm$, the equipartition field strength is about $2\kG$,
and this might be where the sunspot field is strongest.
If NEMPI was to be responsible for this, we should expect the effective
magnetic pressure to be negative at a depth of about $10\Mm$.
Here, the equipartition field strength is about $3\kG$.
If NEMPI operates at $\meanB_z^{\max}/\Beq\approx0.1$, this would
correspond to $\meanB_z^{\max}\approx300\G$, which appears plausible.
At that depth, the density scale height is also about $10\Mm$.
Thus, if magnetic flux concentrations have a size of 5 density scale
heights, then this would correspond to $50\Mm$ at that depth.
To produce spots higher up, the field would need to be more
concentrated, which would reduce the size by a factor of 3 again.
However, given the many uncertainties, it is impossible to draw any
further conclusions until NEMPI has been studied under more realistic
conditions relevant to the Sun.

\section{Conclusions}

Using DNS, ILES and MFS in a wide range of parameters
we have demonstrated that an initially uniform vertical weak magnetic field
in strongly stratified MHD turbulence with large scale separation
results in the formation of circular magnetic spots of equipartition
and super-equipartition field strengths.
Although we have confirmed that the normalized horizontal wavenumber of
magnetic flux concentrations is $k_\perp H_\rho\approx0.8$, as found
earlier for horizontal imposed field
\citep{KBKMR13}, it is now clear that in the nonlinear regime
smaller values can be attained.
This happens in a fashion reminiscent of an inverse cascade or inverse
transfer\footnote{In both cases, the transfer is {\em nonlocal} in
wavenumber space.
It is therefore more appropriate to use the term inverse transfer instead
of inverse cascade.} in helically forced turbulence \citep{B01}.
In the present case, this inverse transfer is found both in
MFS and in DNS.
This property helps explaining the possibility of larger length scales
separating different flux concentrations.

The study of axisymmetric MFS helps understanding the dependence
of NEMPI on the parameters $\betastar$ and $\betap$,
which determine the parameterization of the effective
magnetic pressure, $\Peff(\beta)$.
It was always clear that changes in those parameters can significantly
change the functional form of $\Peff(\beta)$, and yet the resulting
growth rate of NEMPI was found to depend mainly on the value of $\betastar$.
We now see, however, that the shape of the resulting solutions still
depends on the value of $\betap$, in addition to a dependence on $\betastar$.
In fact, smaller values of $\betap$ as well as larger values of $\betastar$
both result in longer structures.
This is important background information in attempts to find flux concentrations
in DNS, where the domain might not always be tall enough.
As a rule of thumb, we can now say that the domain is deep enough if the
resulting large-scale magnetic field is below 1\% of the equipartition value.
This is confirmed by \Figs{pBprof_comp}{pBprof_comp_Rm} as well as
\Figs{fig:pBprof_comp_iles}{fig:pBprof_comp_iles_alt},
where all runs with $\meanB_z^{\rm max}/\Beq(z)\leq0.01$ at $z=z_{\rm bot}$
reach $\meanB_z^{\rm max}/\Beq(z)=O(1)$ at $z=z_{\rm top}$, provided the
domain is also high enough.
A limited extent at the top appears to be less critical than at the bottom,
because NEMPI still develops in almost the same way as before.

It is important to emphasize that the formation of magnetic flux
concentrations is equally well possible at large Mach numbers.
This is important in view of applications to the Sun, where in the
upper layers $\Ma\approx0.5$ can be expected.
Nevertheless, our present investigations have not yet been able to
address the question whether sunspots can really form through NEMPI.
For that, we would need to abandon the assumption of isothermality.
Nevertheless, we expect the basic feature of downflows along flux tubes
to persist also in that case.
It is the associated inflow from the side that keeps the tube concentrated.
Such flows have indeed been seen in local helioseismology \citep{ZKS10}.
Those authors also find an additional outflow higher in the photosphere
that is known as the Evershed flow.

We expect that the downflow in the tube plays an important role in
an unstably stratified layer, such as in the Sun, where it brings
low entropy material to deeper layers, lowering therefore the
effective temperature in the magnetic tubes.
Future work should hopefully be able to demonstrate that in detail.
The conceptual difference between NEMPI and other mechanisms may not
always be very clear.
However, by using an isothermal layer, we can be sure that
convection is not operating.
Thus, the phenomenon of flux segregation found by \cite{TWBP98}
would not work.
Conversely, however, NEMPI might well be a viable explanation for
this phenomenon too.
Whether the concept of flux expulsion can really serve as an
alternative paradigm is unclear, because it is difficult
to draw any quantitative predictions from it.
In particular, flux expulsion does not make any reference to
turbulent pressure or its suppression.
Instead, the source of free energy is more directly potential
energy which can be tapped through the superadiabatic gradient
in convection.
By contrast, the source of free energy for NEMPI is turbulent energy.
The other possibility discussed above is the network of downdrafts
associated with the supergranulation layer \citep{SN12}.
This mechanism is not easily disentangled from NEMPI, because both
imply flux concentrations in downdrafts.
However, in an isothermal layer, we can be sure that supergranulation
flows are absent, so NEMPI is the only known mechanism able to explain
the flux concentrations shown in the present paper.

\begin{acknowledgements}
This work was supported in part by
the European Research Council under the
AstroDyn Research Project No.\ 227952,
by the Swedish Research Council under the project grants
2012-5797 and 621-2011-5076 (AB), by the European Research Council under
the Atmospheric Research Project No.\
227915, and by a grant from the Government of the Russian Federation under
contract No. 11.G34.31.0048 (NK, IR).
We acknowledge the allocation of computing resources provided by the
Swedish National Allocations Committee at the Center for
Parallel Computers at the Royal Institute of Technology in
Stockholm and the National Supercomputer Centers in Link\"oping, the High
Performance Computing Center North in Ume\aa,
and the Nordic High Performance Computing Center in Reykjavik.
Part of this work used the {\sc Nirvana} code version 3.3,
developed by Udo Ziegler at the Leibniz-Institut
f{\"u}r Astrophysik Potsdam (AIP).
\end{acknowledgements}


\end{document}